%% file: main.tex
\documentclass{LMCS}
\usepackage{amssymb,amsmath,stmaryrd,eepic,epic}
\usepackage{gastex}
\usepackage{epsfig}
\usepackage{enumerate,hyperref}

\newcounter{compressEnum}

{}

\newcommand{\equi}{\approx}

\newcommand{\li}{l_0}
\newcommand{\Pref}{{\sf Prefs}}
\newcommand{\Play}{{\sf Plays}}
\newcommand{\K}{{\sf K}}
\newcommand{\Last}{{\sf Last}}

\newcommand{\Knw}{{\sf Knw}}

\newcommand{\ov}{\overline}

\def\sg{\mathrel[\joinrel\mathrel[}
\def\sd{\mathrel]\joinrel\mathrel]}

\def\abs#1{\ensuremath{\lvert #1\rvert}}

 \renewcommand{\S}{{\mathcal{S}}}
 \newcommand{\Epsilon}{{\mathcal{E}}}
 \renewcommand{\L}{{\mathcal{L}}}
\newcommand{\pre}{\mathsf{pre}}

\newcommand{\outcome}{\mathsf{outcome}}

\newcommand{\Outcome}{{\mathsf{Outcome}}}
\newcommand{\trans}{{\Delta}}

\newcommand{\real}{{\mathbb R}}

\newcommand{\sem}[1]{\sg \mathrel{#1} \sd}

\newcommand{\arup}[1]{\lceil \mathrel {#1} \rceil}

\newcommand{\bigarup}[1]{\big\lceil \! \mathrel{#1}\!\! \big\rceil}

\newcommand{\Bigarup}[1]{\Big\lceil \! \mathrel{#1}\!\! \Big\rceil}

\newcommand{\arupL}{\lceil}

\newcommand{\arupR}{\rceil}

\newcommand{\downc}[1]{{#1\!\!\downarrow}}

\renewcommand{\l}{{\ell}}

\newcommand{\tuple}[1]{\langle #1 \rangle}

\newcommand{\Post}{\mathsf{Post}}

\newcommand{\CPre}{\mathsf{CPre}}

\newcommand{\Obs}{{\mathcal{O}}}

\renewcommand{\obs}{o}

\newcommand{\target}{{\mathcal T}}

\newcommand{\last}{\mathsf{Last}}
\newcommand{\DAG}{{\sc dag}}

\newcommand{\sink}{{\sf sink}}

\newcommand{\straa}{\alpha}
\newcommand{\strab}{\beta}
\newcommand{\dist}{{\mathcal D}}
\newcommand{\Supp}{{\sf Supp}}

\newcommand{\Straa}{{\mathcal A}}
\newcommand{\Strab}{{\mathcal B}}

\newcommand{\rank}{\mathsf{Rank}}
\newcommand{\dest}[2]{\Post^H_{#2}(#1)} 

\newcommand{\cone}{\mathsf{Cone}}
\newcommand{\as}{\mathsf{AS}}
\newcommand{\Reach}{\mathsf{Reach}}
\newcommand{\Buchi}{\mathsf{Buchi}}
\newcommand{\coBuchi}{\mathsf{coBuchi}}
\newcommand{\Parity}{\mathsf{Parity}}
\newcommand{\allow}{\mathsf{Allow}}
\newcommand{\supp}{\Supp}
\newcommand{\post}{\Post}
\newcommand{\Safe}{\mathsf{Safe}}
\newcommand{\transg}{\trans}
\newcommand{\transh}{\trans_H}
\newcommand{\apre}{\mathsf{Apre}}
\newcommand{\spre}{\mathsf{Spre}}
\newcommand{\Prb}{\mathrm{Pr}}
\newcommand{\set}[1]{\{\: #1 \:\}}
\newcommand{\distr}{\dist}
\newcommand{\Inf}{\mathsf{Inf}}
\newcommand{\Nats}{\mathbb{N}}

\newcommand{\CharacFormula}[1]{\mathsf{{\mu}Form}(#1)}
\def\anti{{\mathcal A}}

\def\doi{3 (3:4) 2007}
\lmcsheading%
{\doi}
{1--23}
{}
{}
{Jan.~\phantom{0}3, 2007}
{Jul.~27, 2007}
{}

\begin{document}
 
\title[Alg. for $\omega$-Regular Games]{Algorithms for Omega-Regular Games with Imperfect Information\rsuper*}

\author[K. Chatterjee]{Krishnendu Chatterjee\rsuper a}	
\address{{\lsuper a}EECS, University of California at Berkeley, U.S.A.}	
\email{c\_krish@eecs.berkeley.edu}  
\thanks{
This research was supported in part by
the NSF grants CCR-0225610 and CCR-0234690\rsuper{a,c},
by the SNSF under the Indo-Swiss Joint Research Programme\rsuper{b,c},
and by the FRFC project ``Centre F\'ed\'er\'e en V\'erification" 
funded by the FNRS under grant 2.4530.02\rsuper{b,d}.}

\author[L.Doyen]{Laurent Doyen\rsuper b}	
\address{{\lsuper b}CCS, \'{E}cole Polytechnique F\'ed\'erale de Lausanne, Switzerland}	
\email{laurent.doyen@epfl.ch}  

\author[T.~A.~Henzinger]{Thomas A. Henzinger\rsuper c}	
\address{{\lsuper c}CCS, \'{E}cole Polytechnique F\'ed\'erale de Lausanne, Switzerland and EECS, University of California at Berkeley, U.S.A.}
\email{tah@epfl.ch}  

\author[J.-F.~Raskin]{Jean-Fran\c{c}ois Raskin\rsuper d}	
\address{{\lsuper d}CS, Universit\'{e} Libre de Bruxelles, Belgium}	
\email{jraskin@ulb.ac.be}  

\titlecomment{{\lsuper *}A preliminary version of this paper appeared in 
the \emph{Proceedings of the International Conference for Computer Science Logic} (CSL), Lecture Notes in 
Computer Science 4207, Springer, 2006, pp. 287-302.}

\begin{abstract}
We study observation-based strategies for two-player turn-based games
on graphs with omega-regular objectives.  An observation-based
strategy relies on imperfect information about the history of a play,
namely, on the past sequence of observations.  Such games occur in the
synthesis of a controller that does not see the private state of the
plant.  Our main results are twofold.  First, we give a fixed-point
algorithm for computing the set of states from which a player can win
with a deterministic observation-based strategy for any omega-regular
objective.  The fixed point is computed in the lattice of antichains
of state sets.  This algorithm has the advantages of being directed by
the objective and of avoiding an explicit subset construction on the
game graph.  Second, we give an algorithm for computing the set of
states from which a player can win with probability 1 with a
randomized observation-based strategy for a B\"uchi objective.  This
set is of interest because in the absence of perfect information,
randomized strategies are more powerful than deterministic ones.  We
show that our algorithms are optimal by proving matching lower bounds.
\end{abstract}

\maketitle

\vfill\eject

\section{Introduction}

Two-player games on graphs play an important role in computer science.
In particular, the {\em controller synthesis} problem asks, given a
model for a plant, to construct a model for a controller such that the
behaviors resulting from the parallel composition of the two
models respects a given specification (e.g., are included in an
$\omega$-regular set).  Controllers can be synthesized as winning
strategies in a game graph whose vertices represent the plant states,
and whose players represent the plant and the controller
\cite{RamadgeWonham,PnueliR89}.  Other applications of game graphs
include realizability and compatibility checking, where the players
represent parallel processes of a system, or its environment
\cite{AbadiLamportWolper,DillBook,InterfaceAutomata}.

Most results about two-player games played on graphs make the
hypothesis of {\em perfect information}.  In this setting, the
controller knows, during its interaction with the plant, the
exact state of the plant.  In practice, this hypothesis is often not
reasonable.  For example, in the context of hybrid systems, the
controller acquires information about the state of the plant
using sensors with finite precision, which return imperfect
information about the state.  Similarly, if the
players represent individual processes, then a process has only access
to the public variables of the other processes, not to their private
variables \cite{Reif84,AHK02}.

Two-player games of {\em imperfect information} are considerably more
complicated than games of perfect information.  First, decision
problems for imperfect-information games usually lie in higher
complexity classes than their perfect-information counter-parts
\cite{Reif84,kupferman-synthesis,AHK02}.  The algorithmic difference is
often exponential, due to a subset construction that, similar to the
determinization of finite automata, turns an imperfect-information
game into an equivalent perfect-information game.  Second, because of
the determinization, no symbolic algorithms are known to solve 
imperfect-information games.  This is in contrast to the
perfect-information case, where (often) simple and elegant
fixed-point algorithms exist \cite{EJ91,AHM01lics}.  Third, in the context of
imperfect information, deterministic strategies are sometimes 
insufficient.  A game is {\em turn-based\/} if in every state one of the
players chooses a successor state.  In turn-based games of perfect information
the set of winning states coincides with the set of states where the probability 
of winning is~1, and so deterministic strategies
suffice to win (and thus also to win with probability~1).
In contrast, in turn-based games of imperfect information the set of winning states 
is in general a strict subset of the set of states where the probability 
of winning is~1, and so randomized strategies are required to win
with probability~1 (see Example~\ref{ex:example-one}).  
Fourth, winning strategies for
imperfect-information games need memory even for simple objectives
such as safety and reachability (see Example~\ref{ex:example-two}). 
This is again in contrast to the
perfect-information case, where turn-based safety and reachability
games can be won with memoryless strategies.  

The contributions of this paper are twofold.  First, we provide a
symbolic fixed-point algorithm to compute winning states in games of imperfect
information for arbitrary $\omega$-regular objectives.  The novelty is
that our algorithm is symbolic; it does not carry out an explicit
subset construction.  Instead, we compute fixed points on the lattice
of antichains of state sets.  Antichains of state sets can be seen
as a symbolic and compact representation for
$\subseteq$-downward-closed sets of sets of states.\footnote{We
  recently used this symbolic representation of
  $\subseteq$-downward-closed sets of state sets to propose a new
  algorithm for solving the universality problem of nondeterministic
  finite automata. First experiments show a very promising performance;
  see~\cite{CAV06} for details.}  This solution extends our recent
result~\cite{DDR06} from safety objectives to all $\omega$-regular
objectives.  To justify the correctness of the algorithm, we transform
games of imperfect information into games of perfect information
while preserving the existence of winning strategies for every 
objective.  The reduction is only part of the proof, not part of the
algorithm.  For the special case of parity objectives, we obtain a
symbolic {\sc Exptime} algorithm for solving parity games of
imperfect information.  This is optimal, as the reachability problem
for games of imperfect information is known to be {\sc Exptime}-hard~\cite{Reif84}.

Second, we study randomized strategies and winning with probability~1
for imperfect-information games.  To our knowledge, for these games
no algorithms (symbolic or not) are present in the literature.
Following~\cite{AHK98}, we refer to winning with probability~1 as {\em
almost-sure} winning ({\em almost\/} winning, for short), in
contrast to {\em sure} winning with deterministic strategies.  We
provide a symbolic {\sc Exptime} algorithm to compute the set of
almost-winning states for games of imperfect information with B\"uchi
objectives (reachability objectives can be obtained as a special
case, and for safety objectives almost winning and sure winning
coincide).  
Our solution is again justified by a reduction to games of
perfect information.  However, for randomized strategies the
reduction is different, and considerably more complicated.  
We prove our algorithm to be optimal, showing that computing the
almost-winning states for reachability games of imperfect information
is {\sc Exptime}-hard.
The problem of computing the almost-winning states for coB\"uchi 
objectives under imperfect information 
in {\sc Exptime} remains an open problem.

The paper is organized as follows.
Section~2 presents the definitions;
Section~3 gives the algorithm for the case of sure winning with 
deterministic strategies;
Section~4, for the case of almost winning with randomized strategies;
and Section~5 provides the lower bounds.

\medskip\noindent{\em Related work.}
%
In~\cite{PnueliR89}, Pnueli and Rosner study the synthesis
of reactive modules.  In their framework, there is no game graph;
instead, the environment and the objective are specified using an {\sf
LTL} formula.  In~\cite{kupferman-synthesis}, Kupferman and Vardi
extend these results in two directions: they consider ${\sf CTL}^*$
objectives and imperfect information.  Again, no game graph, but a
specification formula is given to the synthesis procedure.  We believe
that our setting, where a game graph is given explicitly, is more
suited to fully and uniformly understand the role of imperfect
information.  For example, Kupferman and Vardi claim that imperfect
information comes at no cost, because if the specification is given as
a ${\sf CTL}$ (or ${\sf CTL}^*$) formula, then the synthesis problem
is complete for {\sc Exptime} (resp. {\sc 2Exptime}), just as in the
perfect-information case.  These hardness results, however, depend on
the fact that the specification is given compactly as a formula.  
In our setting, with an explicit game graph, reachability
games of perfect information are {\sc Ptime}-complete,
whereas reachability games of imperfect information are 
{\sc Exptime}-complete~\cite{Reif84}.
None of the above papers provide symbolic solutions, and none of them
consider randomized strategies.

It is known that for Partially Observable Markov Decision Processes
(POMDPs) with boolean rewards and limit-average objectives the
quantitative analysis (whether the value is greater than a specified
threshold) is {\sc Exptime}-complete~\cite{Littman-thesis}. 
However, almost winning is a qualitative question, and our hardness 
result for almost winning of imperfect-information games does not 
follow from the known results on POMDPs.  We give in Section~5 
a detailed proof of the hardness result of~\cite{Reif84} for sure winning 
of imperfect-information games with reachability objectives, and we show 
that this proof can be extended to almost winning as well.
To the best of our knowledge, this is the first hardness result that
applies to the qualitative analysis of almost winning in 
imperfect-information games.


A class of \emph{semiperfect}-information games, where 
one player has imperfect information and the other player has
perfect information, is studied in~\cite{ChatterjeeH05}.
That class is simpler than the games studied here; 
it can be solved in NP $\cap$ coNP for parity objectives.

\section{Definitions}\label{sec:definitions}

\noindent
  A \emph{game structure} (\emph{of imperfect information}) is a tuple
  $G=\tuple{L,\li,\Sigma,\trans,\Obs,\gamma}$, where $L$ is a finite
  set of states, $\li \in L$ is the initial state, $\Sigma$ is a
  finite alphabet, $\trans \subseteq L \times \Sigma \times L$ is
  a set of labeled transitions, $\Obs$ is a finite set of
  observations, and $\gamma : \Obs \rightarrow 2^L \backslash \emptyset$ maps
  each observation to the set of states that it represents.
  We require the following two properties on~$G$: 
  $(i)$ for all $\l \in L$ and all $\sigma \in \Sigma$, there exists 
  $\l' \in L$
  such that $(\l, \sigma, \l') \in \trans$; and
  $(ii)$ the set $\{\gamma(\obs) \mid  \obs \in \Obs \}$ 
  partitions $L$.
We say that $G$ 
is a game structure of \emph{perfect information}
if $\Obs = L$ and $\gamma(\l) = \{\l\}$ for all $\l \in L$.
We often omit $(\Obs,\gamma)$ in the description of games of perfect 
information. 
For $\sigma \in \Sigma$ and $s \subseteq L$,
let $\Post^G_\sigma(s) = \{\l' \in L \mid \exists \l \in s: (\l,\sigma,\l') \in \trans \}$.

\medskip\noindent{\em Plays.}
In a game structure, in each turn, Player~$1$ chooses
a letter in $\Sigma$, and Player~$2$ resolves
nondeterminism by choosing the successor state. 
A \emph{play} in $G$ is an infinite sequence 
$\pi=\l_0 \sigma_0 \l_1 \ldots \sigma_{n-1} \l_n \sigma_n \ldots$ such that 
$(i)$ $\l_0=\li$, and
$(ii)$ for all $i \geq 0$, we have $(\l_i,\sigma_i,\l_{i+1}) \in \trans$.  
The \emph{prefix up to $\l_n$} of the play $\pi$ is denoted by $\pi(n)$;
its \emph{length} is $\abs{\pi(n)} = n+1$; 
and its \emph{last element} is $\last(\pi(n)) = \l_n$.
The \emph{observation sequence} of $\pi$ is the unique infinite sequence 
$\gamma^{-1}(\pi)=o_0 \sigma_0 o_1 \ldots \sigma_{n-1} o_n \sigma_{n} \ldots$ 
such that for all $i \geq 0$, we have $\l_i \in \gamma(o_i)$. 
Similarly, the \emph{observation sequence} of $\pi(n)$ is the prefix up to 
$o_n$ of~$\gamma^{-1}(\pi)$.
The set of infinite plays in $G$ is denoted $\Play(G)$,
and the set of corresponding finite prefixes is denoted $\Pref(G)$.
A state $\l \in L$ is \emph{reachable} in $G$ if there exists a prefix 
$\rho \in \Pref(G)$ such that $\Last(\rho) = \l$.
For a prefix $\rho \in \Pref(G)$, the \emph{cone} $\cone(\rho)
=\set{\pi \in \Play(G) \mid \rho \mbox{ is a prefix of } \pi}$
is the set of plays that extend $\rho$.
The \emph{knowledge}
associated with a finite observation sequence $\tau=o_0 \sigma_0 o_1
\sigma_1 \ldots \sigma_{n-1} o_n$ is the set $\K(\tau)$ of
states in which a play can be after this sequence of
observations, that is, $\K(\tau)=\{ \Last(\rho) \mid \rho \in \Pref(G)
\textrm{ and } \gamma^{-1}(\rho)=\tau\}$.
%
%
\begin{lem}\label{lem:knowledge}
Let $G=\tuple{L,\li,\Sigma,\trans,\Obs,\gamma}$ 
be a game structure of imperfect information.
For $\sigma \in \Sigma$,  $\l \in L$, and $\rho,\rho' \in \Pref(G)$ with 
$\rho' = \rho \cdot \sigma \cdot \l$, let $\obs_{\l} \in \Obs$ be the 
unique observation such that $\l \in \gamma(\obs_{\l})$.
Then $\K(\gamma^{-1}(\rho')) = \Post^G_\sigma(\K(\gamma^{-1}(\rho))) \cap \gamma(\obs_{\l})$.
\end{lem}

\medskip\noindent{\em Strategies.}
A \emph{deterministic strategy} in $G$ for Player~$1$ is a function 
$\straa:\Pref(G) \to \Sigma$. 
For a finite set $A$, a probability distribution on $A$ is a function
$\kappa: A \to [0,1]$ such that $\sum_{a \in A} \kappa(a) = 1$. 
We denote the set of probability distributions on $A$ by $\dist(A)$. 
Given a distribution $\kappa \in \dist(A)$,
let $\Supp(\kappa) = \{a \in A \mid \kappa(a) > 0\}$ be 
the \emph{support} of $\kappa$.
A \emph{randomized strategy} in $G$ for Player~$1$ is a function 
$\straa:\Pref(G) \to \dist(\Sigma)$. 
A (deterministic or randomized) strategy $\straa$ for Player~$1$ is 
\emph{observation-based} if for all prefixes $\rho,\rho' \in \Pref(G)$, 
if $\gamma^{-1}(\rho)=\gamma^{-1}(\rho')$, then $\straa(\rho)=\straa(\rho')$. 
In the sequel, we are interested in the existence of observation-based
strategies for Player~$1$.
A \emph{deterministic strategy} in $G$ for Player~$2$ is a function 
$\strab:\Pref(G) \times \Sigma \to L$
such that for all $\rho \in \Pref(G)$ and all $\sigma \in \Sigma$, we have
$(\Last(\rho), \sigma, \strab(\rho,\sigma)) \in \trans$.
A \emph{randomized strategy} in $G$ for Player~$2$ is a function 
$\strab:\Pref(G) \times \Sigma \to \dist(L)$
such that for all $\rho \in \Pref(G)$, all $\sigma \in \Sigma$, and all 
$\l \in \Supp(\strab(\rho,\sigma))$, we have 
$(\Last(\rho), \sigma, \l) \in \trans$.
We denote by $\Straa_G$, $\Straa_G^O$, and $\Strab_G$ the set of all 
Player-$1$ strategies, the set of all observation-based Player-$1$ 
strategies, and the set of all Player-$2$ strategies in $G$, respectively.
All results of this paper can be proved also if strategies depend on state 
sequences only, and not on the past moves of a play.

The \emph{outcome} of two deterministic strategies $\straa$ (for
Player~$1$) and $\strab$ (for Player~$2$) in $G$ is the play $\pi=\l_0
\sigma_0 \l_1 \ldots \sigma_{n-1} \l_n \sigma_n \ldots \in \Play(G)$
such that for all $i \geq 0$, we have $\sigma_i = \straa(\pi(i))$ and
$\l_{i+1} = \strab(\pi(i),\sigma_i)$.  
This play is denoted $\outcome(G,\straa,\strab)$.
The \emph{outcome} of two randomized strategies $\straa$ (for
Player~$1$) and $\strab$ (for Player~$2$) in $G$ is the set of plays
$\pi=\l_0 \sigma_0 \l_1 \ldots \sigma_{n-1} \l_n \sigma_n \ldots \in
\Play(G)$ such that for all $i \geq 0$, we have
$\straa(\pi(i))(\sigma_i)>0$ and $\strab(\pi(i),\sigma_i)(\l_{i+1})>0$.
This set is denoted $\outcome(G,\straa,\strab)$.
The \emph{outcome set} of the deterministic (resp.\ randomized)
strategy $\straa$ for Player~$1$ in $G$ is the set
$\Outcome_i(G,\straa)$ of plays $\pi$ such that there exists a
deterministic (resp.\ randomized) strategy $\strab$ for
Player~$2$ with 
$\pi=\outcome(G,\straa,\strab)$ 
(resp.\ $\pi\in\outcome(G,\straa,\strab)$).
The outcome sets for Player~2 are defined symmetrically.

\medskip\noindent{\em Objectives.}
An \emph{objective} for $G$ 
is a set $\phi$ of infinite sequences of  observations and input letters, 
that is, $\phi \subseteq (\Obs\times \Sigma )^\omega$. 
A play $\pi = \l_0 \sigma_0 \l_1 \ldots \sigma_{n-1} \l_n \sigma_n \ldots \in \Play(G)$ 
\emph{satisfies} the objective $\phi$, denoted $\pi \models \phi$, if 
$\gamma^{-1}(\pi) \in \phi$.
Objectives are generally Borel measurable: 
a Borel objective is a Borel set
in the Cantor topology on $(\Obs\times\Sigma)^\omega$~\cite{Kechris}.
We specifically consider reachability, safety,
B\"uchi, coB\"uchi, and parity objectives, all of them Borel measurable.
The parity objectives are a canonical form to express all $\omega$-regular 
objectives~\cite{Thomas97}.
For a play $\pi=\l_0\sigma_0\l_1\dots$, we write 
$\Inf(\pi)$ for the set of observations that 
appear infinitely often in $\gamma^{-1}(\pi)$, that is,
$\Inf(\pi)=\{ o \in \Obs \mid \l_i \in \gamma(o)
\mbox{ for infinitely many } i\mbox{'s} \}$.

\begin{enumerate}[$\bullet$]
\item \emph{Reachability and safety objectives.}
Given a set $\target \subseteq \Obs$ of target observations, the \emph{reachability} objective 
$\Reach(\target)$ requires that an observation in $\target$ be visited at least once, that is, 
$\Reach(\target)=\set{ \l_0 \sigma_0 \l_1 \sigma_1 \ldots 
\in \Play(G) \mid \exists k \geq 0 \cdot \exists \obs \in \target:  \l_k \in \gamma(\obs)}$.
Dually, the \emph{safety} objective $\Safe(\target)$ requires that only observations in $\target$ be 
visited.
Formally,
$\Safe(\target)=\set{ \l_0 \sigma_0 \l_1 \sigma_1 \ldots 
\in \Play(G) \mid \forall k \geq 0 \cdot \exists o \in \target:  \l_k \in \gamma(o)}$.
\item \emph{B\"uchi and coB\"uchi objectives.}
The \emph{B\"uchi} objective $\Buchi(\target)$ 
requires that an observation in $\target$ be visited infinitely often, 
that is,
$\Buchi(\target)=\set{ \pi \mid \Inf(\pi) \cap \target \neq \emptyset}$.
Dually, the \emph{coB\"uchi} objective $\coBuchi(\target)$ requires that 
only observations in $\target$ be visited infinitely often.
Formally, $\coBuchi(\target) =\set{\pi \mid \Inf(\pi) \subseteq \target}$.
\item\emph{Parity objectives.} 
For $d \in \Nats$, let $p:\Obs \to \set{0,1,\ldots,d}$ be a 
\emph{priority function}, 
which maps each observation to a nonnegative integer priority.
The \emph{parity} objective $\Parity(p)$ requires that the minimum priority 
that appears infinitely often be even.
Formally,
$\Parity(p)=\set{\pi \mid \min\set{ p(\obs) \mid \obs \in \Inf(\pi)} 
\mbox{ is even} }$.
\end{enumerate}
Observe that by definition, for all objectives $\phi$, if $\pi \models \phi$ and 
$\gamma^{-1}(\pi)=\gamma^{-1}(\pi')$, then $\pi'\models \phi$.

\medskip\noindent{\em Sure winning and almost winning.}
A strategy $\lambda_i$ for Player~$i$ in $G$ is \emph{sure winning} 
for an objective $\phi$ if for all $\pi \in \Outcome_i(G,\lambda_i)$, 
we have $\pi \models \phi$.
Given a game structure $G$ and a state $\l$ of 
$G$, we write $G_{\l}$ for the game structure that 
results from $G$ by changing the initial state to $\l$, 
that is, if $G=\tuple{L,\li,\Sigma,\trans,\Obs,\gamma}$,
then $G_{\l}=\tuple{L,\l,\Sigma,\trans,\Obs,\gamma}$.
An \emph{event} is a measurable set of plays, and 
given strategies $\straa$ and $\strab$ for the two players, 
the probabilities of events are uniquely defined~\cite{Var85}. 
For a Borel objective~$\phi$,
we denote by $\Prb_{\l}^{\straa,\strab}(\phi)$ 
the probability that $\phi$ is satisfied in the game $G_{\l}$
given the strategies $\straa$ and $\strab$.
A strategy $\straa$ for Player~$1$ in $G$ is \emph{almost winning} for 
the objective $\phi$ if for all randomized strategies $\strab$ for 
Player~$2$, we have $\Prb_{\li}^{\straa,\strab}(\phi)=1$.
The set of \emph{sure-winning} (resp.\ \emph{almost-winning}) \emph{states} 
of a game structure $G$ for the objective 
$\phi$ is the set of states $\l$ such that Player~$1$ has a deterministic 
sure-winning (resp.\ randomized almost-winning) 
observation-based  strategy in $G_{\l}$  for the objective $\phi$.

\begin{thm}[Determinacy] {\rm\cite{Mar75}}
\label{thrm:boreldeterminacy}
For all perfect-information game structures $G$
and all Borel objectives $\phi$,
either there exists a deterministic sure-winning strategy for Player~$1$ for 
the objective $\phi$, or there exists a deterministic sure-winning strategy 
for Player~$2$ for the complementary objective $\Play(G) \setminus \phi$. 
\end{thm}

Notice that deterministic strategies suffice for \emph{sure} winning a game:
given a randomized strategy $\straa$ for Player~$1$, let $\straa^D$
be the deterministic strategy such that for all $\rho \in \Pref(G)$, 
the strategy 
$\straa^D(\rho)$ chooses an input letter from $\supp(\straa(\rho))$. 
Then $\Outcome_1(G,\straa^D) \subseteq \Outcome_1(G,\straa)$, and
thus, if $\straa$ is sure winning, then so is~$\straa^D$.
The result also holds for observation-based strategies.
However, for almost winning, randomized strategies are more powerful 
than deterministic strategies as shown by Example~\ref{ex:example-one}.

\begin{exa}\label{ex:example-one}
Consider the game structure shown in \figurename~\ref{figure-example1}.
The observations $o_1,o_2,o_3,o_4$ are such that 
$\gamma(o_1)=\{ \l_1 \}$, $\gamma(o_2)=\{ \l_2,\l'_2 \}$, 
$\gamma(o_3)=\{ \l_3,\l'_3 \}$, and $\gamma(o_4)=\{ \l_4\}$.
The transitions are shown as labeled edges in the figure, and the 
initial state is $\l_1$.
The objective of Player~$1$ is $\Reach(\{o_4\})$,
to reach state $\l_4$.
We argue that the game is not sure winning for Player~$1$.
Let $\straa$ be any deterministic strategy for Player~$1$.
Consider the deterministic strategy $\strab$ for Player~$2$ as follows: 
for all $\rho \in \Pref(G)$
such that $\Last(\rho) \in \gamma(o_2)$, if $\straa(\rho)=a$, then 
in the previous round $\strab$ chooses the state $\l_2$, and 
if $\straa(\rho)=b$, then 
in the previous round $\strab$ chooses the state $\l'_2$.
Given $\straa$ and~$\strab$, the play $\outcome(G,\straa,\strab)$ 
never reaches $\l_4$.  Similarly, Player~$2$ has no sure winning strategy for
the dual objective $\Safe(\{o_1,o_2,o_3\})$. Hence the game is not determined.
However, the game $G$ is almost winning for Player~$1$.
Consider the randomized strategy that plays $a$ and $b$ uniformly at 
random at all states. 
Every time the game visits observation $o_2$, for any strategy for Player~$2$,
the game visits $\l_3$ and $\l'_3$ with probability $\frac{1}{2}$, and hence
also reaches $\l_4$ with probability $\frac{1}{2}$. 
It follows that against all Player~$2$ strategies the play 
eventually reaches $\l_4$ with probability~1.
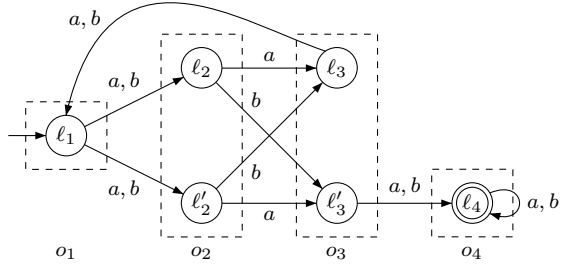
\begin{figure}[t]
   \begin{center}
      \input{figure-example.tex}
   \end{center}
  \caption{Game structure $G$.}
  \label{figure-example1}
\end{figure}
\end{exa}

\paragraph{Spoiling strategies.} To spoil a strategy of Player~$1$ (for sure-winning),
Player~$2$ does not need the full memory of the history of the play, he only needs
counting strategies. We say that a deterministic strategy $\strab: \Pref(G) \times \Sigma \to L$ for Player~$2$ 
is \emph{counting} if for all prefixes $\rho, \rho' \in \Pref(G)$ such that $\abs{\rho} = \abs{\rho'}$
and $\last(\rho) = \last(\rho')$, and for all $\sigma \in \Sigma$, we have $\strab(\rho,\sigma) = \strab(\rho',\sigma)$.
Let $\Strab_G^c$ be the set of counting strategies for Player~$2$. 
The memory needed by a counting strategy is only the number of turns that have been played. This type of 
strategy is sufficient to spoil the non-winning strategies of Player~$1$.

\begin{prop}\label{prop:counting-suffices}
Let $G$ be a game structure of imperfect information and $\phi$ be an 
objective. 
There exists an observation-based strategy $\straa^o \in \Straa_G^O$
such that for all $\strab \in \Strab_G$ we have $\outcome(G,\straa^o, \strab) 
\in \phi$
if and only if there exists an observation-based strategy 
$\straa^o \in \Straa_G^O$ 
such that for all counting strategies $\strab^c \in \Strab_G^c$ 
we have $\outcome(G,\straa^o, \strab^c) \in \phi$.
\end{prop}

\proof
We prove the equivalent statement that:
$\forall \straa^o \in \Straa_G^o \cdot \exists \strab \in \Strab_G: \outcome(G,\straa^o, \strab) \not\in \phi$
iff 
$\forall \straa^o \in \Straa_G^o \cdot \exists \strab^c \in \Strab_G^c: \outcome(G,\straa^o, \strab^c) \not\in \phi$.
The right implication $(\leftarrow)$ is trivial. For the left implication $(\rightarrow)$, let $\straa^o \in \Straa_G^o$
be an arbitrary observation-based strategy for Player~$1$ in $G$. Let $\strab \in \Strab_G$ be a strategy for Player~$2$
such that $\outcome(G,\straa^o, \strab) \not\in \phi$. 
Let $\outcome(G,\straa^o, \strab) = \l_0 \sigma_0 \l_1 \dots \sigma_{n-1} \l_n \sigma_n \dots$ and 
define a counting strategy $\strab^c$ for Player~$2$ such that $\forall \rho \in \Pref(G) \cdot \forall \sigma \in \Sigma:$
if $\last(\rho) = \l_{n-1}$ and $\sigma = \sigma_{n-1}$ for $n= \abs{\rho}$, 
then $\strab^c(\rho,\sigma) = \l_n$, and otherwise $\strab^c(\rho,\sigma)$ is fixed arbitrarily 
in the set $\Post^G_{\sigma}(\last(\rho))$.
Clearly, $\strab^c$ is a counting strategy and we have $\outcome(G,\straa^o, \strab) = \outcome(G,\straa^o, \strab^c)$
and thus $\outcome(G,\straa^o, \strab^c) \not\in \phi$.
\qed

\noindent{\em Remarks.} 
First, the hypothesis that the observations form a partition of
the state space can be weakened to a covering of the state space, where
observations can overlap~\cite{DDR06}. In that case, Player~$2$ chooses both the 
next state of the game $\l$ and the next observation~$\obs$
such that $\l \in \gamma(\obs)$. The definitions related to plays, strategies,
and objectives are adapted accordingly. 
Such a game structure $G$ with overlapping observations can be encoded by
an equivalent game structure $G'$ of imperfect information, whose state space
is the set of pairs $(\l,\obs)$ such that  $\l \in \gamma(\obs)$. 
The set of labeled
transitions $\trans'$ of $G'$ is defined by $\Delta' = \set{((\l,\obs),\sigma,(\l',\obs')) \mid (\l,\sigma, \l') \in\trans}$
and $\gamma'^{-1}(\l,\obs)=\obs$. 
The games $G$ and $G'$ are equivalent in the sense that
for every Borel objective $\phi$, 
there exists a sure (resp.\ almost) winning strategy for Player~$i$
in $G$ for $\phi$ if and only if there exists such a 
winning strategy for Player~$i$ in $G'$ for $\phi$.

Second, it is essential that the objective is expressed in terms of the 
observations. 
Indeed, the games of imperfect information with a nonobservable winning 
condition are more complicated to solve. 
For instance, the universality problem for B\"uchi automata
can be reduced to such games, but the construction that we
propose in Section~\ref{sec:sure-winning} cannot be used. More involved 
constructions \`a la Safra are needed~\cite{Saf88}.

\section{Sure Winning}\label{sec:sure-winning}

First, we show that a game structure $G$ of imperfect information can
be encoded by a game structure $G^{\K}$ of perfect information such
that for every objective $\phi$, there exists a deterministic
observation-based sure-winning strategy for Player~$1$ in $G$ for
$\phi$ if and only if there exists a deterministic sure-winning
strategy for Player~$1$ in $G^{\K}$ for $\phi$.
We obtain $G^{\K}$ using a subset construction similar to Reif's construction
for safety objectives~\cite{Reif84}. 
Each state in $G^{\K}$ is a set of states of $G$ which represents 
the knowledge of Player~$1$. 
In the worst case, the size of $G^{\K}$ is exponentially larger than 
the size of $G$. 
Second, we present a fixed-point algorithm based on antichains of set 
of states~\cite{DDR06}, whose correctness relies on the subset construction,
but avoids the explicit construction of $G^{\K}$.

\subsection{Subset construction for sure winning}\label{sec:subset-construction}

\noindent{\em Subset construction.} 
Given a game structure of imperfect information 
$G=\tuple{L,\li,\Sigma,\trans,\Obs,\gamma}$,
we define the \emph{knowledge-based subset construction} 
of $G$ as the following game structure of perfect information:
\begin{center}
$G^{\K} = \tuple{\L, \{\li\}, \Sigma, \trans^{\K}}$,
\end{center}
where $\L = 2^L \backslash \{\emptyset\}$, and 
$(s_1, \sigma, s_2) \in \trans^{\K}$ iff there exists an observation
$\obs \in \Obs$ such that
$s_2 = \Post^G_\sigma(s_1) \cap \gamma(\obs)$ and $s_2 \neq \emptyset$. 
Notice that for all $s \in \L$ and all $\sigma \in \Sigma$, there exists
a set $s' \in \L$ such that $(s,\sigma,s') \in \trans^{\K}$.

A (deterministic or randomized) strategy in $G^{\K}$ is called a 
\emph{knowledge-based} strategy.
To distinguish between a general strategy in $G$, an observation-based 
strategy in $G$, and a knowledge-based strategy in $G^{\K}$, 
we often use the notations $\straa, \straa^o$, and $\straa^{\K}$, respectively.

\begin{lem}\label{lem:reachable-visible}
For all sets $s \in \L$ that are reachable in $G^{\K}$, 
and all observations $\obs \in \Obs$,
either $s \subseteq \gamma(\obs)$ or $s \cap \gamma(\obs) = \emptyset$.
\end{lem}
\proof
First, the property holds for $s = \{\li\}$, the initial state in $G^{\K}$ as it is a singleton. 
Second, we show that the property holds for any successor $s'$ of any state $s$ in $G^{\K}$. 
Assume that $(s,\sigma,s') \in \trans^{\K}$. 
Then we know that $s' = \Post^G_\sigma(s) \cap \gamma(\obs)$ for some $\obs \in \Obs$.
Hence, $s' \subseteq \gamma(\obs)$ and $s' \cap \gamma(\obs') = \emptyset$ for all $\obs' \neq \obs$ since
the set $\{\gamma(\obs) \mid \obs \in\Obs\}$ partitions $L$.
\qed

Abusing the notation, for a play 
$\pi=s_0 \sigma_0 s_1 \ldots \sigma_{n-1} s_n \sigma_n \ldots \in
\Play(G^{\K})$ 
we define its \emph{observation sequence} as the infinite sequence 
$\gamma^{-1}(\pi)=o_0 \sigma_0 o_1 \ldots \sigma_{n-1} o_n \sigma_{n} \ldots$ 
of observations such that for all $i\geq 0$, 
we have $s_i \subseteq \gamma(o_i)$.
This sequence is unique by Lemma~\ref{lem:reachable-visible}. 
The play $\pi$ \emph{satisfies} an objective 
$\phi \subseteq (\Obs\times \Sigma)^\omega$ if 
$\gamma^{-1}(\pi) \in \phi$.



The correctness of the subset construction $G^{\K}$ is established by the following
two lemmas which generalize the result of~\cite{Reif84} for safety objective 
to any kind of objective. For Lemma~\ref{theo:winning-then-winningK}, the proof
of~\cite{Reif84} is not sufficient, since violation of a safety objective can be witnessed
by a finite prefix of play, while general objectives need an infinite witness. 

\begin{lem}\label{theo:winningK-then-winning}
If Player~$1$ has a deterministic sure-winning strategy in $G^{\K}$ for an objective $\phi$,
then he has a deterministic observation-based sure-winning strategy in $G$ for~$\phi$.
\end{lem}
\proof
Let $\straa^{\K}$ be a deterministic sure-winning strategy for Player~$1$ in $G^{\K}$ with the objective $\phi$.
Define $\straa^o$ a strategy for Player~$1$ in $G$ as follows: 
for every $\rho \in \Pref(G)$, let $\straa^o(\rho) = \straa^{\K}(\rho^{\K})$
where $\rho^{\K}$ is defined from $\rho= \l_0 \sigma_0 \l_1 \dots \sigma_{n-1} \l_n$ 
by $\rho^{\K} = s_0 \sigma_0 s_1 \dots \sigma_{n-1} s_n$ 
where $s_i = \K(\gamma^{-1}(\l_0 \sigma_0 \l_1 \dots \sigma_{i-1} \l_i))$ for each $0 \leq i \leq n$.
Clearly, $\straa^o$ is a deterministic observation-based strategy as 
$\gamma^{-1}(\rho) = \gamma^{-1}(\rho')$ implies $\rho^{\K} = \rho'^{\K}$.

By contradiction, assume that $\straa^o$ is not a sure-winning strategy for Player~$1$ in $G$ with the objective $\phi$. 
Then there exists a play $\pi \in \Outcome_1(G,\straa^o)$ such that $\pi \not\models \phi$.
Let $\pi = \l_0 \sigma_0 \l_1 \sigma_1 \dots$ and consider the infinite sequence $\pi^{\K} = s_0 \sigma_0 s_1 \sigma_1 \dots$
where $s_i = \K(\gamma^{-1}(\pi(i)))$ for each $i \geq 0$. We show that $\pi^{\K} \in \Outcome_1(G^{\K},\straa^{\K})$.
First, we have $s_0 = \K(\gamma^{-1}(\pi(0))) = \K(\gamma^{-1}(\l_0)) = \{\l_0\}$. Second, for any $i \geq 0$, 
we have $s_i = \K(\gamma^{-1}(\pi(i)))$ and by Lemma~\ref{lem:knowledge} we have
$s_{i+1} = \Post^G_{\sigma_i}(s_i) \cap \gamma(\obs)$
where $\obs$ is such that $\l_{i+1} \in \gamma(\obs)$ and so $(s_i, \sigma_i, s_{i+1}) \in \trans^{\K}$. Third, by definition of $\straa^o$,
we have $\sigma_i = \straa^o(\pi(i)) = \straa^{\K}(\pi^{\K}(i))$. This entails that $\pi^{\K} \in \Outcome_1(G^{\K},\straa^{\K})$.

Now, observe that trivially $\l_i \in \K(\gamma^{-1}(\pi(i)))$ for any $i \geq 0$, that is $\l_i \in s_i$ and so 
$s_i \cap \gamma(\obs_i) \neq \emptyset$  where $\obs_i$ is the unique observation such that 
$\l_i \in \gamma(\obs_i)$. 
Hence, by Lemma~\ref{lem:reachable-visible}, 
we have $s_i \subseteq \gamma(\obs_i)$.
Consequently, $\gamma^{-1}(\pi^{\K}) = \gamma^{-1}(\pi)$ and thus $\pi^{\K} \not\models \phi$ which contradicts the fact that
$\straa^{\K}$ is a sure-winning strategy for Player~$1$ in $G^{\K}$ with the objective $\phi$. Therefore, $\straa^o$ 
is a sure-winning strategy for Player~$1$ in $G$ with the objective $\phi$.
\qed

\begin{lem}\label{theo:winning-then-winningK}
If Player~$1$ has a deterministic observation-based sure-winning strategy in $G$ for 
an objective $\phi$, then Player~$1$ has a deterministic sure-winning strategy in $G^{\K}$ 
for $\phi$.
\end{lem}

\proof
First, it is easy to show by induction that for every finite prefix of play 
$\rho^{\K} = s_0 \sigma_0 s_1 \dots \sigma_{n-1} s_n$ in $\Pref(G^{\K})$, there 
exists a prefix of play $\rho = \l_0 \sigma_0 \l_1 \dots \sigma_{n-1} \l_n$
in $\Pref(G)$ that \emph{generates} $\rho^{\K}$, that is such that 
$s_i = \K(\gamma^{-1}(\l_0 \sigma_0 \l_1 \dots \sigma_{i-1} \l_i))$ 
for each $0 \leq i \leq n$; and for all such prefix of play $\rho'$ that 
generates $\rho^{\K}$, we have $\gamma^{-1}(\rho) = \gamma^{-1}(\rho')$ 
(by Lemma~\ref{lem:knowledge}).

Now, let $\straa^o$ be a deterministic observation-based sure-winning strategy for 
Player~$1$ in $G$ that is sure-winning for $\phi$. We construct a deterministic
strategy $\straa^{\K}$ for Player~$1$ in $G^{\K}$ as follows: for every $\rho^{\K} \in \Pref(G^{\K})$,
let $\straa^{\K}(\rho^{\K}) = \straa^o(\rho)$ where $\rho$ generates $\rho^{\K}$.
By the above remark, $\straa^{\K}$ is well-defined (it is independent of the choice
of $\rho$ since $\straa^o$ is observation-based). 

By contradiction, assume that $\straa^{\K}$ is not sure-winning for Player~$1$ 
in $G^{\K}$ with objective $\phi$. Then, there exists a play 
$\pi^{\K} \in \Outcome_1(G^{\K}, \straa^{\K})$ with $\pi^{\K} \not\models \phi$.

We construct the \DAG\/ $D=\tuple{V,E}$ where $V = \{(\l,i) \mid \l \in \Last(\pi^{\K}(i)) \}$
and $E = \{((\l,i),(\l',i+1)) \mid (\l,\sigma_i,\l') \in \Delta\}$. By definition 
of $G^{\K}$, for all $i \geq 0$, we have $\Last(\pi^{\K}(i)) \neq \emptyset$ 
and for all $\l \in \Last(\pi^{\K}(i))$, there is a path in $D$ from $(\l_0,0)$ 
to $(\l,i)$. Therefore, $V$ is infinite and by K\"onig's Lemma, there exists an
infinite path $(\l_0,0) (\l_1,1) \dots$ in $D$ and thus a play 
$\pi = \l_0 \sigma_0 \l_1 \sigma_1 \dots$ in $G$ such that $\pi \in \Outcome_1(G, \straa^o)$
and $\pi \not\models \phi$. This is in contradiction with the assumption that 
$\straa^o$ is sure-winning in $G$ for $\phi$. Hence $\straa^{\K}$ is sure-winning 
for Player~$1$ in $G^{\K}$ with objective $\phi$.
\qed

\noindent 
Lemma~\ref{theo:winningK-then-winning} and 
Lemma~\ref{theo:winning-then-winningK} yield 
Theorem~\ref{coro:winningK-iff-winning}.

\begin{thm}[Sure-winning reduction]\label{coro:winningK-iff-winning}
Player~$1$ has a deterministic 
observation-based sure-winning strategy in a game structure 
$G$ of imperfect information for an objective $\phi$
if and only if 
Player~$1$ has a deterministic sure-winning strategy in 
the game structure $G^{\K}$ of perfect information for~$\phi$.
\end{thm}

\subsection{Two interpretations of the $\mu$-calculus}\label{sec:interpretations-mu-calculus}

From the results of Section~\ref{sec:subset-construction}, we can
solve a game $G$ of imperfect information with objective $\phi$ by
constructing the knowledge-based subset construction $G^{\K}$ 
and solving the resulting game of perfect information for the objective
$\phi$ using standard methods. For the important class of
$\omega$-regular objectives, there exists a fixed-point theory ---the
$\mu$-calculus--- for this purpose~\cite{AHM01lics}.  When run on
$G^{\K}$, these fixed-point algorithms compute sets of sets of states
of the game $G$.  An important property of those sets is that they are
\emph{downward closed} with respect to set inclusion: if Player~$1$ has a
deterministic strategy to win the game $G$ when her knowledge is a
set $s$, then she also has a deterministic strategy to win the game
when her knowledge is $s'$ with $s' \subseteq s$. And thus, if $s$ is
a sure-winning state of $G^{\K}$, then so is $s'$.  Based on this
property, we devise a new algorithm for solving games of perfect
information.

An \emph{antichain} of nonempty sets of states is a set $q \subseteq
2^L \setminus \{\emptyset\}$ such that for all $s,s' \in q$, we have $s
\not\subset s'$.  Let $\anti$ be the set of antichains of nonempty
subsets of $L$, and consider the following partial order on $\anti$: for
all $q,q' \in \anti$, let $q \sqsubseteq q'$ iff $\forall s \in q \cdot
\exists s' \in q': s \subseteq s'$. For $q \subseteq 2^L$, 
define the set of \emph{maximal} elements of $q$ by
$\arup{q} = \{s \in q \mid s \neq \emptyset \mbox{ and } \forall s' \in q
: s\not\subset s' \}$. Clearly, $\arup{q}$ is an antichain.  The least upper
bound of $q, q' \in \anti$ is $q \sqcup q' = \arup{\{s \mid s \in q \mbox{ or }
  s \in q'\}}$, and their greatest lower bound is $q \sqcap q' =
\arup{\{s \cap s' \mid s \in q \mbox{ and } s' \in q'\}}$.  The definition of
these two operators extends naturally to sets of antichains, and the
greatest element of $\anti$ is $\top = \{L\}$ and the least element is
$\bot = \emptyset$. The partially ordered set $\tuple{\anti, \sqsubseteq,
  \sqcup, \sqcap, \top, \bot}$ forms a complete lattice.  We view
antichains of state sets as a symbolic representation of
$\subseteq$-downward-closed sets of state sets.

A \emph{game lattice} is a complete lattice $V$ together with a \emph{predecessor operator}
$\CPre: V \to V$.
Given a game structure $G=\tuple{L,\li,\Sigma,\trans,\Obs,\gamma}$ 
of imperfect information, and its
knowledge-based subset construction 
$G^{\K} = \tuple{\L, \{\li\}, \Sigma, \trans^{\K}}$,
we consider two game lattices: the \emph{lattice of subsets} 
$\tuple{\S, \subseteq, \cup, \cap, \L, \emptyset}$, where $\S = 2^{\L}$
and $\CPre: \S \to \S$ is defined by
$
{\sf CPre}(q)= \{ s \in \L \mid \exists \sigma \in \Sigma \cdot \forall s'\in\L:
\textrm{ if } (s,\sigma,s') \in \trans^{\K}, \textrm{ then } s'\in q \};
$
and the \emph{lattice of antichains} $\tuple{\anti, \sqsubseteq, \sqcup, \sqcap, \{L\}, \emptyset}$,
with the operator $\arup{\CPre}: \anti \to \anti$ defined by 
$
\arup{\CPre}(q) = \arupL \{s \in \L \mid \exists \sigma \in \Sigma \cdot \forall \obs \in \Obs \cdot \exists s' \in q:
{\sf Post}_{\sigma}(s) \cap   \gamma(\obs) \subseteq s'\} \arupR.
$
 
The \emph{$\mu$-calculus formulas} are generated by the grammar
$$ \varphi ::= \obs \mid x \mid \varphi \lor \varphi \mid \varphi \land \varphi \mid \pre(\varphi) \mid \mu x.\varphi \mid \nu x.\varphi$$
for atomic propositions $\obs \in \Obs$ and variables $x$. 
We can define $\lnot \obs$ as a shortcut for $\bigvee_{\obs' \in \Obs \backslash \{\obs\}} \obs'$.
A variable is \emph{free} in a formula $\varphi$ if it is not in the scope of a 
quantifier $\mu x$ or $\nu x$. A formula $\varphi$ is \emph{closed} if it contains no free variable.
Given a game lattice $V$, a \emph{valuation} $\Epsilon$ for the variables 
is a function that maps every variable $x$ to an element in $V$. For $q \in V$, we write $\Epsilon[x \mapsto q]$ for the valuation 
that agrees with $\Epsilon$ on all variables, except that~$x$ is mapped to $q$. Given a game lattice $V$ and a valuation $\Epsilon$,
each $\mu$-calculus  formula $\varphi$ specifies an element $\sem{\varphi}^V_\Epsilon$ of $V$, which is defined inductively by the
equations shown in the two tables below.
If $\varphi$ is a closed formula, then $\sem{\varphi}^{V} = \sem{\varphi}^{V}_\Epsilon$ for any valuation $\Epsilon$.
The following theorem recalls that perfect-information games can be solved 
by evaluating fixed-point formulas in the lattice of subsets.

\begin{figure}[t]
\begin{center}
{\small
\begin{minipage}{8cm}
\begin{tabular}{|l|}
\hline
\multicolumn{1}{|c|}{{\Large \strut} Lattice of subsets}\\
\hline
{\Large \strut} $\sem{\obs}^{\S}_\Epsilon = \{s \in \L \mid s \subseteq \gamma(\obs)\}$ \\
{\Large \strut} $\sem{x}^{\S}_\Epsilon = \Epsilon(x)$ \\
{\Large \strut} $\sem{\varphi_1 \Big\{\! \begin{array}{c} \lor \\[-2pt] \land \end{array}\!\Big\} \varphi_2}^{\S}_\Epsilon = 
\sem{\varphi_1}^{\S}_\Epsilon \Big\{\! \begin{array}{c} \cup \\[-2pt] \cap \end{array}\!\Big\} \sem{\varphi_2}^{\S}_\Epsilon$ \\
{\Large \strut} $\sem{\pre(\varphi)}^{\S}_\Epsilon = \CPre(\sem{\varphi}^{\S}_\Epsilon)$ \\
{\large \strut} $\sem{\Big\{\! \begin{array}{c} \mu \\[-2pt] \nu \end{array}\!\Big\}x. \varphi}^{\S}_\Epsilon = 
\Big\{\! \begin{array}{c} \cap \\[-2pt] \cup \end{array}\!\Big\} \{q  \mid q =
\sem{\varphi}^{\S}_{\Epsilon[x \mapsto q]} \}$ \\
\hline
\end{tabular}
\end{minipage}
\begin{minipage}{8cm}
\begin{tabular}{|l|}
\hline
\multicolumn{1}{|c|}{{\Large \strut} Lattice of antichains} \\
\hline
{\Large \strut} $\sem{\obs}^{\anti}_\Epsilon = \{\gamma(\obs)\}$ \\
{\Large \strut} $\sem{x}^{\anti}_\Epsilon = \Epsilon(x)$ \\
{\Large \strut} $\sem{\varphi_1 \Big\{\! \begin{array}{c} \lor \\[-2pt] \land \end{array}\!\Big\} \varphi_2}^{\anti}_\Epsilon = 
\sem{\varphi_1}^{\anti}_\Epsilon \Big\{\! \begin{array}{c} \sqcup \\[-2pt] \sqcap \end{array}\!\Big\} \sem{\varphi_2}^{\anti}_\Epsilon$ \\
{\Large \strut} $\sem{\pre(\varphi)}^{\anti}_\Epsilon = \arup{\CPre}(\sem{\varphi}^{\anti}_\Epsilon)$ \\
{\Large \strut} $\sem{\Big\{\! \begin{array}{c} \mu \\[-2pt] \nu \end{array}\!\Big\}x. \varphi}^{\anti}_\Epsilon = 
\Big\{\! \begin{array}{c} \sqcap \\[-2pt] \sqcup \end{array}\!\Big\} \{q  \mid q =
\sem{\varphi}^{\anti}_{\Epsilon[x \mapsto q]} \}$ \\
\hline
\end{tabular}
\end{minipage}
}
\end{center}
\end{figure}

\begin{thm}[Symbolic solution of perfect-information games]
{\rm\cite{AHM01lics}}
\label{theo:omega-regular-mu-calculus}
For every $\omega$-regular objective $\phi$,
there exists a closed $\mu$-calculus formula $\CharacFormula{\phi}$, 
called the \emph{characteristic formula} of $\phi$,
such that for all game structures $G$ of perfect information, 
the set of sure-winning states of $G$ for $\phi$ is 
$\sem{\CharacFormula{\phi}}^{\S}$.
\end{thm}

\noindent{\em Downward closure.} 
Given a set $q \in \S$, the \emph{downward closure} of $q$ is the set
$\downc{q} = \{s \in \L \mid \exists s' \in q: s \subseteq s'\}$.
Observe that in particular, for all $q \in \S$, we have $\emptyset
\not\in \downc{q}$ and $\downc{\arup{q}} = \downc{q}$.  The sets
$\downc{q}$, for $q \in \S$, are the \emph{downward-closed} sets.  A
valuation $\Epsilon$ for the variables in the lattice $\S$ of subsets is 
\emph{downward closed} if
every variable $x$ is mapped to a downward-closed set, that is,
$\Epsilon(x) = \downc{\Epsilon(x)}$.

\begin{lem}\label{lem:antichains-and-downward-closed-sets}
All downward-closed sets $q,q' \in \S$ satisfy
$\arup{q \cap q'} = \arup{q} \sqcap \arup{q'}$ and
$\arup{q \cup q'} = \arup{q} \sqcup \arup{q'}$.
\end{lem}

\begin{lem}\label{lem:mu-calculus-and-downward-closed-sets}
For all $\mu$-calculus formulas $\varphi$
and all downward-closed valuations $\Epsilon$ in the lattice of subsets, 
the set $\sem{\varphi}^{\S}_{\Epsilon}$ is downward closed.
\end{lem}
\proof
We prove this lemma by induction on the structure of $\varphi$.
\begin{enumerate}[$\bullet$]
\item if $\varphi \equiv \obs$ for $\obs \in \Obs$. It is immediate to show that $\sem{\varphi}^{\S}_{\Epsilon} = \downc{\sem{\varphi}^{\S}_{\Epsilon}\,}$.
\item if $\varphi \equiv x$ for a variable $x$. We have $\sem{\varphi}^{\S}_{\Epsilon} = \Epsilon(x)$ which is downward closed by hypothesis.
\item if $\varphi \equiv \varphi_1 \Big\{ \begin{array}{c} \lor \\[-2pt] \land \end{array}\Big\} \,\varphi_2$ and both $\sem{\varphi_1}^{\S}_{\Epsilon}$
and $\sem{\varphi_2}^{\S}_{\Epsilon}$ are downward closed. Then we have $\sem{\varphi}^{\S}_{\Epsilon} \,= \,
\sem{\varphi_1}^{\S}_\Epsilon \Big\{\begin{array}{c} \cup \\[-2pt] \cap \end{array}\Big\} \sem{\varphi_2}^{\S}_\Epsilon$ and the result
follows from the fact that union and intersection of downward closed sets are downward closed.
\item if $\varphi \equiv pre(\varphi_1)$ and $\sem{\varphi_1}^{\S}_{\Epsilon}$ is downward closed. We show that $\sem{\varphi}^{\S}_{\Epsilon}$
is downward closed. Let $s_1 \in \sem{\varphi}^{\S}_{\Epsilon}$ and let $s_2 \in \L$ such that $s_2 \subseteq s_1$. Let us show that 
$s_2 \in \sem{\varphi}^{\S}_{\Epsilon}$. By definition of $\CPre$, since $s_1 \in \sem{pre(\varphi_1)}^{\S}_{\Epsilon}$, there exists $\sigma \in \Sigma$ such that 
for any $s'_1$, if $(s_1,\sigma,s'_1) \in \trans^{\K}$ then $s'_1\in \sem{\varphi_1}^{\S}_{\Epsilon}$. Consider any $s'_2$ such that 
$(s_2,\sigma,s'_2) \in \trans^{\K}$. 
According to the definition of $G^{\K}$, 
we have $s'_2 = 
\Post_{\sigma}(s_2) \cap \gamma(\obs) \neq \emptyset$ for some $\obs \in \Obs$. Now, let $s'_1 = \Post_{\sigma}(s_1) \cap \gamma(\obs)$.
Since $s_2 \subseteq s_1$, we have $s'_2 \subseteq s'_1$ and thus $s'_1 \neq \emptyset$. Therefore $(s_1,\sigma,s'_1) \in \trans^{\K}$
and so $s'_1\in \sem{\varphi_1}^{\S}_{\Epsilon}$. As the latter set is downward closed, we also have $s'_2\in \sem{\varphi_1}^{\S}_{\Epsilon}$
and thus $s_2 \in \sem{pre(\varphi_1)}^{\S}_{\Epsilon}$.
\item if $\varphi \equiv \nu x.\varphi_1$ and $\sem{\varphi_1}^{\S}_{\Epsilon'}$ is downward closed for any downward closed valuation $\Epsilon'$.
By Tarski's theorem, $\sem{\varphi}^{\S}_{\Epsilon}$ is one of the set in the infinite sequence $q_0, q_1, \dots$ defined
by $q_0 = \L$ and for every $i \geq 1$, $q_i = \sem{\varphi}^{\S}_{\Epsilon[x \mapsto q_{i-1}]}$.
Since $q_0$ is downward closed, every $q_i$ ($i \geq 1$) is also downward closed by the induction hypothesis.
\item if $\varphi \equiv \mu x.\varphi_1$ and $\sem{\varphi_1}^{\S}_{\Epsilon'}$ is downward closed for any downward closed valuation $\Epsilon'$. 
The proof is similar to the previous case.\qed
\end{enumerate}

\begin{lem}\label{theo:mu-calculus-equiv}
For all $\mu$-calculus formulas $\varphi$,
and all downward-closed valuations $\Epsilon$ in the lattice of subsets, 
we have 
$\bigarup{\sem{\varphi}^{\S}_{\Epsilon}} = \sem{\varphi}^{\anti}_{\arup{\Epsilon}}$,
where $\arup{\Epsilon}$ is a valuation in the lattice of antichains defined by 
$\arup{\Epsilon}(x) = \arup{\Epsilon(x)}$ for all variables $x$.
\end{lem}
\proof
We prove this by induction on the structure of $\varphi$.
\begin{enumerate}[$\bullet$]
\item if $\varphi \equiv \obs$ for $\obs \in \Obs$. The claim is immediate.
\item if $\varphi \equiv x$ for a variable $x$. We have  $\bigarup{\sem{\varphi}^{\S}_{\Epsilon}} = \arup{\Epsilon(x)} = \sem{\varphi}^{\anti}_{\arup{\Epsilon}}$.
\item if $\varphi \equiv \varphi_1 \Big\{\begin{array}{c} \lor \\[-2pt] \land \end{array}\Big\} \,\varphi_2$ and both 
$\bigarup{\sem{\varphi_1}^{\S}_{\Epsilon}} = \sem{\varphi_1}^{\anti}_{\arup{\Epsilon}}$ and $\bigarup{\sem{\varphi_2}^{\S}_{\Epsilon}} = \sem{\varphi_2}^{\anti}_{\arup{\Epsilon}}$.
Using Lemma~\ref{lem:antichains-and-downward-closed-sets} and Lemma~\ref{lem:mu-calculus-and-downward-closed-sets}, we have successively:

$\begin{array}{l}
\bigarup{\sem{\varphi}^{\S}_{\Epsilon}} = 
\Bigarup{\sem{\varphi_1}^{\S}_\Epsilon \Big\{ \begin{array}{c} \cup \\[-2pt] \cap \end{array}\Big\}\sem{\varphi_2}^{\S}_\Epsilon} = \\
\bigarup{\sem{\varphi_1}^{\S}_\Epsilon} \Big\{ \begin{array}{c} \sqcup \\[-2pt] \sqcap \end{array}\Big\} \bigarup{\sem{\varphi_2}^{\S}_\Epsilon} =
\sem{\varphi_1}^{\anti}_{\arup{\Epsilon}} \Big\{ \begin{array}{c} \sqcup \\[-2pt] \sqcap \end{array}\Big\} \sem{\varphi_2}^{\anti}_{\arup{\Epsilon}}.
\end{array}
$

\item if $\varphi \equiv pre(\varphi_1)$ and $\bigarup{\sem{\varphi_1}^{\S}_{\Epsilon}} = \sem{\varphi_1}^{\anti}_{\arup{\Epsilon}}$.
	\begin{enumerate}[(1)]
	\item We prove the inclusion $\sem{\varphi}^{\anti}_{\arup{\Epsilon}} \subseteq \bigarup{\sem{\varphi}^{\S}_{\Epsilon}}$.
	First, let $s \in \arup{\CPre}(\sem{\varphi_1}^{\anti}_{\arup{\Epsilon}})$. 
	We know that there exists $\sigma \in \Sigma$ such that 
	$\forall o \in \Obs \cdot \exists s'' \in \sem{\varphi_1}^{\anti}_{\arup{\Epsilon}}: 
	{\sf Post}_{\sigma}(s) \cap\gamma(o) \subseteq s''$. Since $\bigarup{\sem{\varphi_1}^{\S}_{\Epsilon}} = 
	\sem{\varphi_1}^{\anti}_{\arup{\Epsilon}}$ (induction hypothesis), it is clear that for such $\sigma$, 
	if $(s,\sigma,s') \in \trans^{\K}$, then there exists $s'' \in \bigarup{\sem{\varphi_1}^{\S}_{\Epsilon}}$
	such that $s' \subseteq s''$.
	And since $\sem{\varphi_1}^{\S}_{\Epsilon}$ is downward closed (by Lemma~\ref{lem:mu-calculus-and-downward-closed-sets}) 
	we have $s' \in \sem{\varphi_1}^{\S}_{\Epsilon}$, so that $s \in \CPre(\sem{\varphi_1}^{\S}_{\Epsilon})$ 
	(and thus $s \in \sem{\varphi}^{\S}_{\Epsilon}$).

	Second, we show that $s$ is maximal in $\sem{\varphi}^{\S}_{\Epsilon}$. By contradiction, 
	assume that there exists $s_1 \in \sem{\varphi}^{\S}_{\Epsilon}$ with $s \subset s_1$.
	Then, by the same argument as in the first part of the proof of the inclusion 
	$\bigarup{\sem{\varphi}^{\S}_{\Epsilon}} \subseteq \sem{\varphi}^{\anti}_{\arup{\Epsilon}}$ below, we have 
	that $s_1$ satisfies the definition of $\arup{\CPre}(\sem{\varphi_1}^{\anti}_{\arup{\Epsilon}})$
	up to the operator $\arup{\cdot}$. This means that $s$ is not maximal in  
	$\arup{\CPre}(\sem{\varphi_1}^{\anti}_{\arup{\Epsilon}})$,	a contradiction.

	\item We prove the inclusion $\bigarup{\sem{\varphi}^{\S}_{\Epsilon}} \subseteq \sem{\varphi}^{\anti}_{\arup{\Epsilon}}$.
	This is trivial if $\bigarup{\sem{\varphi}^{\S}_{\Epsilon}} = \emptyset$. Otherwise, let us first show
	that $\sem{\varphi_1}^{\anti}_{\arup{\Epsilon}} \neq \emptyset$. Let $s \in \bigarup{\sem{\varphi}^{\S}_{\Epsilon}}$.
	Then, there exists $\sigma \in \Sigma$ such that for any $s'$, 
	if $(s,\sigma,s') \in \trans^{\K}$ then $s'\in \sem{\varphi_1}^{\S}_{\Epsilon}$. 
	Since the transition relation of $G$ is total and the observations partition the state space, 
	we have $\Post_{\sigma}(s) \cap \gamma(\obs) \neq \emptyset$ for some $\obs \in \Obs$. Therefore,
	$\sem{\varphi_1}^{\S}_{\Epsilon}$ is nonempty and so is $\sem{\varphi_1}^{\anti}_{\arup{\Epsilon}}$.

	Now, we proceed with the proof of inclusion. Let $s \in \bigarup{\sem{\varphi}^{\S}_{\Epsilon}}$, and 
	let $\sigma \in  \Sigma$ such that such that for any $s'$, 
	if $(s,\sigma,s') \in \trans^{\K}$ then $s'\in \sem{\varphi_1}^{\S}_{\Epsilon}$. 
	Let us show that $s \in \arup{\CPre}(\sem{\varphi_1}^{\anti}_{\arup{\Epsilon}})$. First, consider an arbitrary
	observation $\obs \in \Obs$ and let $s' =  \Post_{\sigma}(s) \cap \gamma(\obs)$. We must show
	that there exists $s'' \in \sem{\varphi_1}^{\anti}_{\arup{\Epsilon}}$ such that $s' \subseteq s''$.
	This is obvious if $s' = \emptyset$ since $\sem{\varphi_1}^{\anti}_{\arup{\Epsilon}}$ is nonempty.
	Otherwise, by the definition of $G^{\K}$,
 	we have $(s,\sigma,s') \in \trans^{\K}$
	and therefore $s'\in \sem{\varphi_1}^{\S}_{\Epsilon}$. Since 
	$\sem{\varphi_1}^{\anti}_{\arup{\Epsilon}} = \bigarup{\sem{\varphi_1}^{\S}_{\Epsilon}}$ (induction hypothesis),
	there exists $s'' \in \sem{\varphi_1}^{\anti}_{\arup{\Epsilon}}$ such that $s' \subseteq s''$, and thus
	$s'$ satisfies the definition of $\arup{\CPre}(\sem{\varphi_1}^{\anti}_{\arup{\Epsilon}})$
	up to the operator $\arup{\cdot}$.

	Second, let us show that $s$  is maximal in $\arup{\CPre}(\sem{\varphi_1}^{\anti}_{\arup{\Epsilon}})$. By contradiction, 
	assume that there exists $s_1 \in \arup{\CPre}(\sem{\varphi_1}^{\anti}_{\arup{\Epsilon}})$ with $s \subset s_1$.
	Then, by the same argument as in the first part of the proof of the inclusion 
	$\sem{\varphi}^{\anti}_{\arup{\Epsilon}} \subseteq \bigarup{\sem{\varphi}^{\S}_{\Epsilon}}$, we have 
	$s_1 \in \sem{\varphi}^{\S}_{\Epsilon}$. This implies that $s \not\in \bigarup{\sem{\varphi}^{\S}_{\Epsilon}}$, 
	a contradiction.
	\end{enumerate}

\item if $\varphi \equiv \nu x.\varphi_1$ and $\bigarup{\sem{\varphi_1}^{\S}_{\Epsilon'}} = \sem{\varphi_1}^{\anti}_{\arup{\Epsilon'}}$
for any downward closed valuation $\Epsilon'$. 
By Tarski's theorem, $\sem{\varphi}^{\S}_{\Epsilon}$ is one of the set in the infinite sequence $q_0, q_1, \dots$ defined
by $q_0 = \L$ and for every $i \geq 1$, $q_i = \sem{\varphi}^{\S}_{\Epsilon[x \mapsto q_{i-1}]}$; and similarly,
$\sem{\varphi}^{\anti}_{\arup{\Epsilon}}$ is one of the set in the infinite sequence $q'_0, q'_1, \dots$ defined
by $q'_0 = \{L\}$ and for all $i \geq 1$, $q_i = \sem{\varphi}^{\anti}_{\arup{\Epsilon}[x \mapsto q'_{i-1}]}$.
Observe that $q'_0 = \arup{q_0}$. By induction, assume that $q'_{i-1} = \arup{q_{i-1}}$ for some $i \geq 1$.
Then $q'_{i} = \arup{q_{i}}$ as $\arup{\Epsilon}[x \mapsto q'_{i-1}] = \bigarup{\Epsilon[x \mapsto q_{i-1}]}$.


\item if $\varphi \equiv \mu x.\varphi_1$ and $\bigarup{\sem{\varphi_1}^{\S}_{\Epsilon'}} = \sem{\varphi_1}^{\anti}_{\arup{\Epsilon'}}$
for any downward closed valuation $\Epsilon'$. The proof is similar to the previous case.\qed
\end{enumerate}

\noindent
Consider a game structure $G$ of imperfect information
and a parity objective~$\phi$.
From Theorems~\ref{coro:winningK-iff-winning} and 
\ref{theo:omega-regular-mu-calculus} and
Lemma~\ref{theo:mu-calculus-equiv}, we can decide the existence of a
deterministic observation-based sure-winning strategy for Player~$1$
in $G$ for $\phi$
without explicitly constructing the knowledge-based subset construction 
$G^{\K}$, by instead evaluating a fixed-point formula in the lattice of 
antichains.

\begin{thm}[Symbolic solution of imperfect-information games]
Let $G$ be a game structure of imperfect information with initial state $\li$. 
For every $\omega$-regular objective~$\phi$, 
Player~$1$ has a deterministic observation-based 
strategy in $G$ for $\phi$ if and only if $\{\li\} \sqsubseteq \sem{\CharacFormula{\phi}}^{\anti}$.
\end{thm}

\begin{cor}\label{coro:pairtyEXPTIME}
Let $G$ be a game structure of imperfect information, 
let $p$ be a priority function, and let $\ell$ be a state of~$G$.
Whether $\ell$ is a sure-winning state in $G$ for the parity objective 
$\Parity(p)$ can be decided in {\sc Exptime}.
\end{cor}

\noindent
Corollary~\ref{coro:pairtyEXPTIME} is proved as follows: 
for a parity objective $\phi$, an equivalent 
$\mu$-calculus formula $\varphi$ can be obtained, where the 
size and the fixed-point quantifier alternations of
$\varphi$ is polynomial in $\phi$.
Thus given $G$ and $\phi$, we can evaluate $\varphi$ in $G^{\K}$
in {\sc Exptime}.

\section{Almost Winning}\label{sec:randomized}

Given a game structure $G$ of imperfect information, we first construct 
a game structure $H$ in which the knowledge of Player 1 is made explicit.
However, the construction is different from the one used for sure winning.
Then, we establish certain equivalences between randomized strategies in $G$ and $H$.
Finally, we show how the reduction can be used to obtain a symbolic 
{\sc Exptime} algorithm for computing almost-winning states in $G$ for B\"uchi 
objectives.
An {\sc Exptime} algorithm for almost winning for coB\"uchi
objectives under imperfect information remains unknown.

\subsection{Subset construction for almost winning}\label{subsec:subset-almost}

\noindent 
Given a game structure of imperfect information
$G=\tuple{L,\li,\Sigma,\transg,\Obs,\gamma}$, 
we construct the game structure
$H=\Knw(G)=\tuple{Q,q_{0},\Sigma,\transh}$ as follows:
$Q =\set{(s,\l)\mid \exists \obs\in \Obs: s \subseteq \gamma(\obs)
\mbox{ and } \l \in s}$;
the initial state is $q_{0}=(\set{\li},\li)$;
the transition relation $\transh \subseteq Q \times \Sigma \times Q$ 
is defined by
$((s,\l),\sigma,(s',\l')) \in \transh$ iff there is an observation 
$\obs \in \Obs$ such that 
$s'=\post_\sigma^G(s) \cap \gamma(o)$ and $(\l,\sigma,\l') \in \transg$.
Intuitively, 
when $H$ is in state $(s,\l)$, it corresponds to $G$ being in state $\l$ and
the knowledge of Player $1$ being~$s$.
%
%
Two states $q=(s,\l)$ and $q'=(s',\l')$ of $H$ are \emph{equivalent}, 
written $q \equi q'$, if $s=s'$, that is when the knowledge of Player 1 is the
same in the two states.
Two prefixes $\rho=q_0 \sigma_0 q_1 \ldots \sigma_{n-1} q_n$ 
and $\rho'=q'_0 \sigma'_0 q'_1 \ldots \sigma'_{n-1} q'_n$ of $H$ 
are \emph{equivalent}, written $\rho \equi \rho$, if
for all $0\leq i \leq n$, we have $q_i\equi q'_i$, and for all 
$0\leq i\leq n-1$, we have $\sigma_i=\sigma'_i$.
Two plays $\pi,\pi'\in \Play(H)$ 
are \emph{equivalent}, written $\pi \equi \pi'$, if
for all $i\geq 0$, we have $\pi(i) \equi \pi'(i)$.
For a state $q \in Q$, we denote by 
$[q]_\equi=\set{q' \in Q \mid q \equi q'}$ the 
$\equi$-equivalence class of $q$. 
We define equivalence classes for prefixes and plays similarly. 
We cannot reuse the results of Section~\ref{sec:sure-winning} to compute
almost-winning states of $G$, as the randomized strategies in $H$ should
not distinguish equivalent states. 

\medskip\noindent{\em Equivalence-preserving strategies and objectives.}
A strategy $\straa$ for Player~$1$ in $H$ is \emph{positional} if it
is independent of the prefix of plays and depends only on the last
state, that is,
for all $\rho,\rho'\in \Pref(H)$ with $\Last(\rho) =\Last(\rho')$,
we have $\straa(\rho) =\straa(\rho')$. 
A positional strategy $\straa$ can be viewed as a function 
$\straa: Q \to \distr(\Sigma)$.
A strategy $\straa$ for Player~$1$ in $H$ is \emph{equivalence-preserving} 
if for all $\rho,\rho' \in \Pref(H)$ with $\rho \equi \rho'$,
we have $\straa(\rho) =\straa(\rho')$.
We denote by $\Straa_H$, $\Straa_H^P$, and $\Straa_H^\equi$ the set of all 
Player-1 strategies, the set of all positional Player-1 strategies, and the 
set of all equivalence-preserving Player-1 strategies in $H$, respectively.
We write $\Straa_H^{\equi(P)}=\Straa_H^\equi \cap \Straa_H^P$ for the set of 
equivalence-preserving positional strategies. 

An objective $\phi$ for $H$ is a subset of $(Q\times \Sigma)^\omega$,
that is, the objective $\phi$ is a set of plays.
The objective $\phi$ is \emph{equivalence-preserving} 
if for all plays $\pi \in \phi$, we have $[\pi]_\equi \subseteq \phi$.

\medskip\noindent{\em Relating prefixes and plays.}
We define a mapping $h:\Pref(G) \to \Pref(H)$ 
that maps prefixes in $G$ to prefixes in $H$ as follows:
given $\rho =\l_0 \sigma_0 \l_1 \sigma_1 \ldots \sigma_{n-1} \l_n$, 
let $h(\rho) =q_0 \sigma_0 q_1 \sigma_1 \ldots \sigma_{n-1} q_n$,
where for all $0\le i\le n$, we have $q_i=(s_i,\l_i)$, and 
for all $0\le i\le n-1$, we have $s_i=\K(\gamma^{-1}(\rho(i)))$.
The following properties hold:
$(i)$ for all $\rho,\rho' \in \Pref(G)$, 
if $\gamma^{-1}(\rho) =\gamma^{-1}(\rho')$, 
then $h(\rho) \equi h(\rho')$;
and 
$(ii)$ for all $\rho,\rho' \in \Pref(H)$, 
if $\rho \equi \rho'$, 
then $\gamma^{-1}(h^{-1}(\rho)) = \gamma^{-1}(h^{-1}(\rho'))$.
The mapping $h:\Play(G)\to\Play(H)$ for plays is defined similarly,
and has similar properties.

\medskip
\noindent{\em Relating strategies for Player~$1$.} 
We define two strategy mappings $\ov{g}:\Straa_H \to \Straa_G$
and $\ov{h}:\Straa_G \to \Straa_H$.
Given a Player-1 strategy $\straa_H$ in $H$, we construct a 
Player-1 strategy $\straa_G =\ov{g}(\straa_H)$ in $G$ as follows: 
for all $\rho \in \Pref(G)$, let 
$\straa_G(\rho) =\straa_H(h(\rho))$.
Similarly, given a Player-1 strategy $\straa_G$ in $G$, 
we construct a Player-1 strategy 
$\straa_H=\ov{h}(\straa_G)$ in $H$ as follows:
for all $\rho \in \Pref(H)$, let 
$\straa_H(\rho) =\straa_G(h^{-1}(\rho))$.
The following properties hold:
$(i)$ for all strategies $\straa_H\in\Straa_H$,
if $\straa_H$ is equivalence-preserving, 
then $\ov{g}(\straa_H)$ is observation-based;
and $(ii)$ for all strategies $\straa_G\in\Straa_G$,
if $\straa_G$ is observation-based, 
then $\ov{h}(\straa_G)$ is equivalence-preserving.

\medskip\noindent{\em Relating strategies for Player~$2$.} 
Observe that for all $q \in Q$, all $\sigma \in \Sigma$, and all $\l' \in L$,
we have $\abs{\set{s' \mid (q,\sigma,q')\in \transh \land q'=(s',\l') }}\leq 1$.
Given a Player-2 strategy $\strab_H$ in $H$, we construct a Player-2 
strategy $\strab_G =\ov{g}(\strab_H)$ as follows: 
for all $\rho \in \Pref(G)$, $\sigma \in \Sigma$, and $\l' \in L$, let 
$\strab_G(\rho,\sigma)(\l') =\strab_H(h(\rho),\sigma)(s',\l')$
if for $s=\K(\gamma^{-1}(\rho))$ and $\l = \last(\rho)$, we have 
$((s,\l), \sigma, (s',\l')) \in \transh$ for some (and then unique) $s'$,
and $\strab_G(\rho,\sigma)(\l') = 0$ otherwise.
Similarly, given a Player-2 strategy $\strab_G$ in $G$, 
we construct a Player-2 strategy 
$\strab_H=\ov{h}(\strab_G)$ in $H$ as follows:
for all $\rho \in \Pref(H)$, all $\sigma\in \Sigma$, 
and all $(s,\l)$, let 
$\strab_H(\rho,\sigma)((s,\l)) =\strab_G(h^{-1}(\rho),\sigma)(\l)$.

\begin{lem}\label{prop:randomized4}
The following assertions hold.
\begin{enumerate}[\em(1)]
\item
 For all $\rho_H \in \Pref(H)$, 
 for every equivalence preserving strategy $\straa_H$, 
 for every strategy $\strab_H$ we have 
\[
\Prb_{q_{0}}^{\straa_H,\strab_H}(\cone(\rho_H)) =
\Prb_{\li}^{\ov{g}(\straa_H),\ov{g}(\strab_H)}(h^{-1}(\cone(\rho_H))).
\]

\item
 For all $\rho_G \in \Pref(G)$, for every observation-based strategy $\straa_G$, 
 for every strategy $\strab_G$ we have 
\[
\Prb_{\li}^{\straa_G,\strab_G}(\cone(\rho_G) ) =
\Prb_{q_{0}}^{\ov{h}(\straa_G),\ov{h}(\strab_G)}(h(\cone(\rho_G))).
\] 
\end{enumerate}
\end{lem}

\proof
The following properties follow from the construction of 
strategies in Section~\ref{subsec:subset-almost}.
\begin{enumerate}[(1)]
\item
For all $\rho_H \in \Pref(H)$, 
for every equivalence preserving strategy $\straa_H$, 
for every strategy $\strab_H$ and for all $\rho_H'$ such that 
$\abs{\rho_H'} = \abs{\rho_H} +1$, 
we have 
$\Prb_{q_{0}}^{\straa_H,\strab_H}\big(\cone(\rho_H') \mid \cone(\rho_H)\big) =
\Prb_{\li}^{\ov{g}(\straa_H),\ov{g}(\strab_H)}\big(h^{-1}(\cone(\rho_H')) \mid h^{-1}(\cone(\rho_H))\big).$

\item
For all $\rho_G \in \Pref(G)$, for every observation-based strategy $\straa_G$, 
for every strategy $\strab_G$ and for all $\rho_G'$ such that 
$\abs{\rho_G'} = \abs{\rho_G} +1$, 
we have 
$\Prb_{\li}^{\straa_G,\strab_G}\big(\cone(\rho_G') \mid \cone(\rho_G)\big) =
\Prb_{q_{0}}^{\ov{h}(\straa_G),\ov{h}(\strab_G)}\big(h(\cone(\rho_G')) \mid 
h(\cone(\rho_G))\big).$ 
\end{enumerate}

The proof for the first part is as follows:
observe that 
\[
\begin{array}{rcl}
\Prb_{q_{0}}^{\straa_H,\strab_H}(\cone(q_{0})) & = &
\Prb_{\li}^{\ov{g}(\straa_H),\ov{g}(\strab_H)}(h^{-1}(\cone(q_0))) \\[2ex]
& = & \Prb_{\li}^{\ov{g}(\straa_H),\ov{g}(\strab_H)}(\cone(\li))=1.
\end{array}
\]
The result follows from the above property and induction. 
The second part follows analogously.
\qed

\begin{thm}[Almost-winning reduction]\label{thrm:randomized1}
Let $G$ be a game structure of imperfect information, and
let $H=\Knw(G)$.
For all Borel objectives $\phi$ for~$G$, 
all observation-based Player-1 strategies $\straa_G$ in $G$, 
and all Player-2 strategies $\strab_G$ in $G$, we have 
$\Prb_{\li}^{\straa_G,\strab_G}(\phi) 
=\Prb_{q_0}^{\ov{h}(\straa_G),\ov{h}(\strab_G)}(h(\phi))$.
Dually, 
for all equivalence-preserving Borel objectives $\phi$ for $H$, 
all equivalence-preserving Player-1 strategies $\straa_H$ in $H$, 
and all Player-2 strategies $\strab_H$ in $H$, we have 
$\Prb_{q_0}^{\straa_H,\strab_H}(\phi) 
=\Prb_{\li}^{\ov{g}(\straa_H),\ov{g}(\strab_H)}(h^{-1}(\phi))$.
\end{thm}
\proof
By the Carathe\'odary unique-extension theorem, a probability measure defined 
on cones has a unique extension to all Borel objectives.
The theorem then follows from Lemma~\ref{prop:randomized4}.
\qed

\noindent
Corollary~\ref{coro:randomized1} follows from Theorem~\ref{thrm:randomized1}.

\begin{cor}\label{coro:randomized1}
For every Borel objective $\Phi_G$ for $G$, we have
\[
\displaystyle
\sup_{\straa_G \in \Straa_G^O} \ \ \inf_{\strab_G \in \Strab_G} \  
\Prb_{\li}^{\straa_G,\strab_G} (\Phi_G)
=
\sup_{\straa_H \in \Straa_H^\equi} \ \ \inf_{\strab_H \in \Strab_H} \  
\Prb_{q_{0}}^{\straa_H,\strab_H} (h(\Phi_G));
\] 
\[
\begin{array}{ll}
\exists \straa_G \in \Straa_G^O. \ \forall \strab_G \in \Strab_G: &   
\Prb_{\li}^{\straa_G,\strab_G}(\Phi_G)  =  1 \\
& \mbox{ iff } 
\exists \straa_H \in \Straa_H^\equi. \ \forall \strab_H \in \Strab_H: \ 
\Prb_{q_{0}}^{\straa_H,\strab_H}(h(\Phi_G))=1.
\end{array}
\]
\end{cor}

\subsection{Almost winning for B\"uchi objectives}
We first illustrate the need of memory and randomization for almost-winning
in imperfect information games with B\"uchi objectives.

\begin{exa}[Memory is needed to almost-win]\label{ex:example-two}
Consider the example of \figurename~\ref{ex-rand}.  
The objective of Player 1 is to reach a state with observation $o_4$.

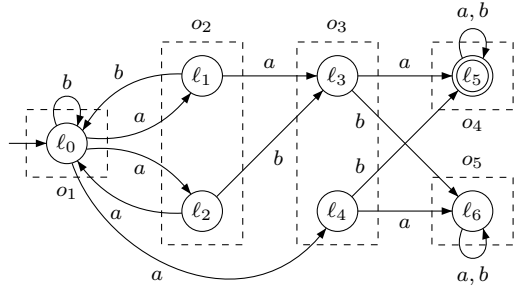
\begin{figure*}[!t]
  \begin{center}
    \input{ex-rand-gastex}
     \caption{Memory and randomization are necessary to almost win the objective $\Buchi(\{o_4\})$.}
    \label{ex-rand}
  \end{center}
\end{figure*}

We show that Player 1 has no observation-based sure-winning strategy in this 
game.
This is because when we fix an observation-based strategy for Player
1, Player 2 has a spoiling strategy to maintain the game into the
states $\{\l_0,\l_1,\l_2\}$. Indeed, at $\l_0$, the only reasonable choice
for Player 1 is to play $a$. Then Player 2 can choose to go either in
$\l_1$ or $\l_2$. In both cases, the observation will be the same for
Player 1. After seeing $o_1 a o_2$,  if the strategy of Player 1 is to play
$a$ then Player 2 chooses $\l_2$, otherwise, if Player 1 strategy is
to play $b$ then Player 2 chooses $\l_1$. This can be repeated and so
Player 2 has a spoiling strategy against any observation-based
strategy of Player 1.

We now show that almost-winning strategies exist for Player~1.
Consider that Player 1 plays an observation-based
randomized strategy $\straa$ as follows: 
after a sequence of observations $\tau$,
  \begin{enumerate}[$\bullet$]
  \item if $\last(\tau)=o_1$, then $\straa(\tau)(a)=1$ and $\straa(\tau)(b)=0$,
  \item if $\last(\tau)=o_2$, then $\straa(\tau)(a)=0.5$ and $\straa(\tau)(b)=0.5$,
  \item if $\tau=\tau'\cdot o_1 \cdot \sigma \cdot o_3$, then $\straa(\tau)(a)=0$ and $\straa(\tau)(b)=1$,
  \item if $\tau=\tau'\cdot o_2 \cdot \sigma \cdot o_3$, then $\straa(\tau)(a)=1$ and $\straa(\tau)(b)=0$,
  \item otherwise take, arbitrarily, $\straa(\tau)(a)=1$ and $\straa(\tau)(b)=0$.
  \end{enumerate}
  
  The strategy $\straa$ is almost-winning against any
  randomized strategy of Player 2. Note that the strategy $\straa$
  uses memory and this is necessary because when receiving observation
  $o_3$, Player 1 has to play $a$ if the previous state satisfied observation 
  $o_1$ and $b$ if the previous state satisfied $o_2$.
\end{exa}

Given a game structure $G$ of imperfect information, 
let $H=\Knw(G)$.
Given a set $\target\subseteq \Obs$ of target observations,
let $B_{\target} =\set{(s,\l) \in Q \mid \exists \obs \in \target:
s \subseteq \gamma(o)}$.
Then $h(\Buchi(\target)) =\Buchi(B_{\target})=\set{\pi_H \in \Play(H) \mid 
\Inf(\pi_H) \cap B_{\target} \neq \emptyset}$.
We first show that almost winning in $H$ for the B\"uchi objective
$\Buchi(B_{\target})$ with respect 
to equivalence-preserving strategies is equivalent to almost winning 
with respect to equivalence-preserving positional strategies. 
Formally, for $B_{\target} \subseteq Q$, let
$
Q_\as^\equi  = \set{q \in Q \mid \exists \straa_H \in \Straa_H^\equi\cdot
\forall \strab_H \in \Strab_H\cdot
\forall q' \in [q]_\equi: \ \Prb_{q'}^{\straa_H,\strab_H}(\Buchi(B_{\target}))=1}$, 
and
%
$
Q_\as^{\equi(P)} = \set{q \in Q \mid \exists \straa_H \in \Straa_H^{\equi(P)}\cdot
\forall \strab_H \in \Strab_H\cdot
\forall q' \in [q]_\equi: \ \Prb_{q'}^{\straa_H,\strab_H}(\Buchi(B_{\target}))=1}.
$
We will prove that $Q_\as^\equi=Q_\as^{\equi(P)}$.
Lemma~\ref{prop:randomized-property} 
follows from the construction of $H$ from $G$,
and yields Lemma~\ref{lemm:randomized-property}.

\begin{lem}\label{prop:randomized-property}
For all $q_1 \in Q$, and all $\sigma \in \Sigma$,
if $(q_1,\sigma,q_1') \in \transh$, then for all $q_2' \in [q_1']_\equi$,
there exists $q_2 \in [q_1]_\equi$ such that $(q_2,\sigma,q_2') \in \transh$.
\end{lem}

\begin{lem}\label{lemm:randomized-property}
Given an equivalence-preserving Player-1 strategy $\straa_H\in\Straa_H$, 
a prefix $\rho\in \Pref(H)$, and a state $q\in Q$,  
if there exists a Player-2 strategy $\strab_H\in\Strab_H$ such that 
$\Prb_q^{\straa_H,\strab_H}(\cone(\rho))>0$, then for every prefix 
$\rho'\in\Pref(H)$ with $\rho\equi\rho'$, there exist a 
Player-2 strategy $\strab'_H\in\Strab_H$ and a state $q' \in [q]_\equi$ 
such that $\Prb_{q'}^{\straa_H,\strab'_H}(\cone(\rho'))>0$.
\end{lem}

\noindent
Observe that 
$
Q \setminus Q_\as^\equi  =  \set{q \in Q \mid \forall \straa_H 
\in \Straa_H^\equi\cdot \exists \strab_H \in \Strab_H\cdot 
\exists q' \in [q]_\equi: 
\Prb_{q'}^{\straa_H,\strab_H}(\Buchi(B_{\target}))<1}.
$
It follows from Lemma~\ref{lemm:randomized-property} that 
if a play starts in $Q_\as^\equi$ and reaches $Q\setminus Q_\as^\equi$ with 
positive probability,
then for all equivalence-preserving strategies for Player~$1$, there is a 
Player~$2$ strategy that ensures that the B\"uchi objective $\Buchi(B_{\target})$ is 
satisfied with probability strictly lower than~1.

\medskip\noindent{\em Notation.}
For a state $q\in Q$ and $Y \subseteq Q$, let 
$\allow(q,Y) =\set{\sigma \in \Sigma\mid \dest{q}{\sigma} \subseteq Y}$.
For a state $q \in Q$ and $Y \subseteq Q$, let
$\allow([q]_\equi,Y)= \bigcap_{q'\in [q]_\equi} \allow(q',Y)$.

\begin{lem}\label{lemm:randomized1}
For all $q\in Q_\as^\equi$, we have 
$\allow([q]_\equi,Q_\as^\equi)\neq \emptyset$.
\end{lem} 
\proof
Assume towards contradiction that there exists $q \in Q_\as^\equi $ such
that $\allow([q]_\equi,Q_\as^\equi)=\emptyset$.
Then for all $\sigma \in \Sigma$ there exists $q' \in [q]_{\equi}$ 
such that $\dest{q'}{\sigma} \cap (Q\setminus Q_\as^\equi)\neq \emptyset$.
Hence for every equivalence preserving strategy $\straa_H$ there exists 
$q'\in [q]_{\equi}$ such that $\straa_H(q')(\sigma)>0$ and 
$\dest{q'}{\sigma} \cap (Q\setminus Q_\as^\equi) \neq \emptyset$.
Hence for every equivalence strategy $\straa_H$ 
there is a state $q' \in [q]_{\equi}$ and a strategy $\strab_H$ for 
Player~$2$ such that $Q \setminus Q_\as^\equi$ is reached with positive probability.
This contradicts that $[q]_\equi \subseteq Q_\as^\equi$.
\qed

\begin{lem}\label{lemm:randomized2}
Given a state $q \in Q_\as^\equi$, 
let $\straa_H\in\Straa_H$ be an equivalence-preserving Player-1 strategy 
such that for all Player-2 strategies $\strab_H\in\Strab_H$ and all 
states $q' \in [q]_\equi$, we have 
$\Prb_{q'}^{\straa_H,\strab_H}(\Buchi(B_{\target}))=1$. 
Let $\rho=q_0\sigma_0 q_1 \ldots \sigma_{n-1} q_n$ be a prefix in $\Pref(H)$ 
such that for all $0 \leq i \leq n$, we have $q_i \in Q_\as^\equi$.
If there is a Player-2 strategy $\strab_H\in\Strab_H$ and a  
state $q' \in [q]_\equi$ such that 
$\Prb_{q'}^{\straa_H,\strab_H}(\cone(\rho))>0$, then  
$\supp(\straa_H(\rho)) \subseteq \allow([q_n]_\equi,Q_\as^\equi)$.
\end{lem}

\proof
Fix an almost-winning strategy $\straa_H$.
Assume towards contradiction for a history 
$\rho_H$ satisfying the conditions of the lemma that there exists  
$\sigma \in \supp(\straa_H(\rho_H))\setminus\allow([q_n],Q_\as^\equi)$.
Then there exists $q_n' \in [q_n]_{\equi}$ such that 
$\dest{q_n'}{\sigma} \cap (Q\setminus Q_\as^\equi) \neq \emptyset$.
Then there exists $\rho_H'$ such that $\rho_H \equi \rho_H'$ and 
$\Last(\rho_H')=q_n'$.
Then by Lemma~\ref{lemm:randomized-property} there exists a strategy 
$\strab_H'$ and $q' \in [q]_\equi$ such that 
$\Prb_{q'}^{\straa_H,\strab_H'}(\cone(\rho_H'))>0$.
Then given $\rho_H'$ and the strategy $\straa_H$ there exists a Player~$2$ 
strategy such that $Q\setminus Q_\as^\equi$ is reached with positive 
probability.
This contradicts that $\straa_H$ is an almost-winning strategy.
\qed

\noindent{\em Notation.}
We inductively define the \emph{ranks} of states in $Q_\as^\equi$ as follows:
let $\rank(0)=B_{\target} \cap Q_\as^\equi$, and for all $j\ge 0$, let 
$
\rank(j+1) =\rank(j)  \cup        
\set{  q \in Q_\as^\equi  \mid 
\exists   \sigma \in \allow([q]_\equi,Q_\as^\equi):
\dest{q}{\sigma} \subseteq 
\rank(j)}.
$
Let $j^*=\min\set{j\ge 0 \mid \rank(j)=\rank(j+1)}$,
and let $Q^*=\rank(j^*)$.
We say that the set $\rank(j+1)\setminus\rank(j)$ contains the 
\emph{states of rank}~$j+1$, for all $j\ge 0$.

\begin{exa}\label{ex:example-three}
Given the game structure $G$ of imperfect information from 
\figurename~\ref{figure-example1}, the  
game structure $H=\Knw(G)$ is shown in 
\figurename~\ref{reduction-ex}.
All states are almost winning for the B\"uchi objective 
$\Buchi(\set{(\{\l_4\},\l_4)})$.
The ranks of the states 
are shown next to the states.
The positional strategy that plays both $a$ and $b$ with equal probability is
almost winning at all states.
For the states $q$ with ranks 1, 3, and~4, 
if the rank of $q$ is $j$, then 
$\post_a^H(q) \subseteq \rank(j-1)$ and
$\post_b^H(q) \subseteq \rank(j-1)$.
For the states with rank~2, 
if $q=(\set{\l_2,\l_2'},\l_2)$, 
then $\post_b^H(q) \subseteq \rank(1)$; and
if $q=(\set{\l_2,\l_2'},\l_2')$, 
then $\post_a^H(q) \subseteq \rank(1)$.  
\begin{figure}[t]
  \begin{center}
    \input{reduction-gastex}
    \caption{Game structure $H = \Knw(G)$ (for $G$ of \figurename~\ref{figure-example1}).}
    \label{reduction-ex}
  \end{center}
\end{figure}
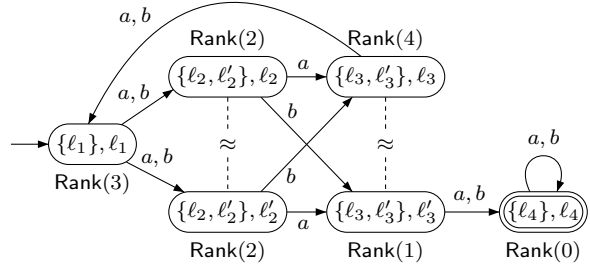
\end{exa}

\begin{lem}\label{lemm:randomized3}
$Q^* = Q_\as^\equi$.
\end{lem}
\proof 
By definition, $Q^* \subseteq Q_\as^\equi$.
We now prove that $Q_\as^\equi \subseteq Q^*$.
Assume towards a contradiction that 
$X=Q_\as^\equi\setminus Q^* \neq \emptyset$.
For all states $q\in X$ and all $\sigma \in \allow([q]_\equi,Q_\as^\equi)$, 
we have $\dest{q}{\sigma} \cap X \neq \emptyset$,
because otherwise $q$ would have been in $Q^*$.
Hence, for all $q \in X$ and all $\sigma \in \allow([q]_\equi,Q_\as^\equi)$, 
there exists a $q' \in X$ such that $(q,\sigma,q') \in \transh$.
Fix a strategy $\strab_H$ for Player~$2$ as follows: 
for a state $q \in X$ and the input letter 
$\sigma \in \allow([q]_\equi,Q_\as^\equi)$, choose
a successor $q' \in X$ such that $(q,\sigma,q') \in \transh$.
Consider a state $q \in X$ and an equivalence-preserving 
almost-winning strategy $\straa_H$ for Player~$1$ from $q$
for the objective $\Buchi(B_{\target})$. 
By Lemma~\ref{lemm:randomized2}, for every prefix $\rho$ 
satisfying the condition of Lemma~\ref{lemm:randomized2}, we have 
$\supp(\straa_H(\rho)) \subseteq \allow([\last(\rho)]_\equi,Q_\as^\equi)$.
It follows that $\Prb_q^{\straa_H,\strab_H}(\Safe(X))=1$.
Since $B_{\target} \cap Q_\as^\equi \subseteq Q^*$, 
it follows that $B_{\target} \cap X =\emptyset$.
Hence  $\Prb_q^{\straa_H,\strab_H}(\Reach(B_{\target}))=0$,
and therefore $\Prb_q^{\straa_H,\strab_H}(\Buchi(B_{\target}))=0$. 
This contradicts the fact that $\straa_H$ is an 
almost-winning strategy.
\qed

\noindent{\em Equivalence-preserving positional strategy.}
Consider the equivalence-preserving positional strategy $\straa_H^p$ 
for Player~1 in $H$, which is defined as follows:
for a state $q \in Q_\as^\equi$, choose all moves in 
$\allow([q]_\equi,Q_\as^\equi)$ uniformly at random.

\begin{lem}\label{lemm:randomized4}
For all states $q \in Q_\as^\equi$ and all Player-2 strategies 
$\strab_H$ in $H$, we have 
$\Prb_q^{\straa_H^p,\strab_H}(\Safe(Q_\as^\equi))=1$ and
$\Prb_q^{\straa_H^p,\strab_H}(\Reach(B_{\target} \cap Q_\as^\equi))=1$.
\end{lem} 

\proof
By Lemma~\ref{lemm:randomized3}, we have $Q^*=Q_\as^\equi$. 
Let $z=|Q^*|$.

\begin{enumerate}[$\bullet$]
\item 
For a state $q \in Q_\as^\equi$, 
we have $\dest{q}{\sigma} \subseteq Q_\as^\equi$
for all $\sigma \in \allow([q]_\equi,Q_\as^\equi)$.
It follows for all states $q \in Q_\as^\equi$ and all strategies 
$\strab_H$ for Player~$2$, we have 
$\Prb_q^{\straa_H^p,\strab_H}(\Safe(Q_\as^\equi))=1$. 
\item

For a state $q \in (\rank(j+1)\setminus \rank(j))$,
there exists $\sigma \in \allow([q]_\equi,Q_\as^\equi)$ such that  
$\dest{q}{\sigma} \subseteq \rank(j) $.
For a set $Y \subseteq Q$, let $\Diamond^j(Y)$ denote the set of 
prefixes that reach $Y$ after at most $j$ steps.
It follows that for all states $q \in \rank(j+1)$
and all strategies $\strab_H$ for Player~2, we have 
\[
\Prb_q^{\straa_H^p,\strab_H}(\Diamond^1(\rank(j)))\geq 
\frac{1}{|\Sigma|}.
\]
Let $B=B_{\target} \cap Q_\as^\equi$.
By induction on the ranks it follows that for all states 
$q \in Q^*$ and all strategies $\strab$ for Player~2: 
\[
\Prb_q^{\straa_H^p,\strab_H}(\Diamond^z(\rank(0))) 
 = 
\Prb_q^{\straa_H^p,\strab_H}(\Diamond^z(B)) 
 \geq 
\Big(\frac{1}{|\Sigma|}\Big)^z =r >0.
\]
For $m>0$, we have
$\Prb_q^{\straa_H^p,\strab_H}(\Diamond^{m\cdot z}(B))
\geq 1- (1-r)^m$.
Thus: 
\[
\Prb_q^{\straa_H^p,\strab_H}(\Reach(B)) =  \lim_{m \to \infty}
\Prb_q^{\straa_H^p,\strab_H}(\Diamond^{m\cdot z}(B)) \\
\geq  \lim_{m \to \infty}
1- (1-r)^m =1. 
\]
\end{enumerate}
The lemma follows.
\qed

Lemma~\ref{lemm:randomized4} implies that, given the Player-1 
strategy $\straa_H^p$, the set $Q_\as^\equi$ is never left, and the states in
$B_{\target} \cap Q_\as^\equi$ are reached with probability~1.
Since this happens for every state in $Q_\as^\equi$, it follows that the
set $B_{\target} \cap Q_\as^\equi$ is visited infinitely often with 
probability~1,
that is, the B\"uchi objective $\Buchi(B_{\target})$ is satisfied with 
probability~1.
This analysis, together with the fact that $[q_{0}]_\equi$ is a singleton and
Corollary~\ref{coro:randomized1}, proves that $Q_\as^\equi=Q_\as^{\equi(P)}$. 
Theorem~\ref{thrm:randomized2} follows.

\begin{thm}[Positional almost winning for B\"uchi objectives
under imperfect information]
\label{thrm:randomized2}
Let $G$ be a game structure of imperfect information, and
let $H=\Knw(G)$.
For all sets $\target$ of observations, 
there exists an observation-based almost-winning strategy 
for Player~1 in $G$ for the objective $\Buchi(\target)$ iff 
there exists an equivalence-preserving positional almost-winning strategy 
for Player~1 in $H$
for the objective $\Buchi(B_{\target})$.
\end{thm}

\noindent{\em Symbolic algorithm.}
We present a symbolic quadratic-time (in the size of $H$) algorithm to 
compute the set $Q_\as^\equi$.
For $Y\subseteq Q$ and $X \subseteq Y$, let 
$
\apre(Y,X) =  \set{q \in Y \mid \exists \sigma \in \allow([q]_\equi,Y): 
	\dest{q}{\sigma} \subseteq X};
\text{ and }
\spre(Y)  =  \set{q\in Y \mid \allow([q]_\equi,Y) \neq \emptyset}. $
Note that $\spre(Y) = \apre(Y,Y)$.
Let 
\[
\phi=\nu Y. \mu X. \big( \apre(Y,X) \lor (B_{\target} \land \spre(Y) \big)
\]
and let $Z=\sem{\phi}$.

\begin{lem}\label{lemm:randomized-symbolic}
$Z=Q_\as^\equi$.
\end{lem}

\noindent{\it Proof\ }[of Lemma~\ref{lemm:randomized-symbolic}].
We prove $Z=Q_\as^\equi$ by proving inclusion in both directions.
We have $Z=\sem{\phi}$ and 
$\phi=\nu Y. \mu X. \big(\apre(Y,X) \lor (B_{\target} \land \spre(Y))\big)$
\begin{enumerate}[(1)]
\item We first show that $Z\subseteq Q_\as^\equi$. 
Since $Z$ is a fixed-point of $\phi$ we have
\[
Z= \sem{\mu X. \big(\apre(Z,X) \lor (B_{\target} \land \spre(Z))\big)}.
\]
We analyze the evaluation of $Z$ as the fixed-point as follows:
let $X_0=\emptyset$ and $X_{i+1}= \apre(Z,X_i) \lor (B_{\target} \land \spre(Z))$.
Observe that since $X_0=\emptyset$ we have $\apre(Z,X_0)=\emptyset$ and 
hence $X_1= B_{\target} \cap \spre(Z) \subseteq B_{\target}$.
Let $j^* =\min\set{i \mid X_{i+1} =X_i}$ and we have $Z=X_{j^*}$.
Consider the equivalence preserving strategy $\straa_H^p$ for Player~1 that 
at a state $q \in Z$ plays all moves 
in $\allow([q]_\equi,Z)$ uniformly at random.
For all $q \in Z$, for all $q' \in [q]_\equi$, and for all $\sigma \in \allow([q]_\equi,Z)$,
we have $\dest{q'}{\sigma} \subseteq Z$.
It follows that for all strategies $\strab_H$ for Player~2 and for all states $q \in Z$ 
we have $\Prb_q^{\straa_H^p,\strab_H}(\Safe(Z))=1$.
Also for a state $q \in (X_{i+1} \setminus X_i) \setminus B_{\target}$ we have 
there exists $\sigma \in \allow([q]_\equi,Z)$ such that $\post^H_\sigma(q) \subseteq
X_i$, i.e., for a state $q\in (X_{i+1} \setminus X_i) \setminus B_{\target}$, given 
$\straa_H^p$ against all strategies $\strab_H$ the next state is in $X_i$
with probability at least $\frac{1}{|\Sigma|}$. 
Arguments similar to Lemma~\ref{lemm:randomized4} establishes that 
$\straa_H^p$ is an almost-winning strategy for all states $q \in Z$.
Hence we have $Z \subseteq Q_\as^\equi$.

\item We now show that $Q_\as^\equi \subseteq Z$.
We first show that $Q_\as^\equi$ satisfies that 
\[
Q_\as^\equi=\sem{\mu X.  \big(\apre(Q_\as^\equi,X) \lor 
(B_{\target} \land \spre(Q_\as^\equi))\big)}.
\]
Observe that $Q_\as^\equi=\spre(Q_\as^\equi)$. 
We now analyze the evaluation of $Q_\as^\equi$ as the fixed-point $Q^*$ 
as shown in Lemma~\ref{lemm:randomized3}.
Let $X_0=\emptyset$, then $\apre(Q_\as^\equi,X_0)=\emptyset$.
Hence $X_1= \apre(Q_\as^\equi) \lor (B_{\target} \land \spre(Q_\as^\equi)) 
=B_{\target} \land Q_\as^\equi= \rank(0)$ 
(as defined before Lemma~\ref{lemm:randomized3}). 
By the definition of $\rank(j+1)$ from $\rank(j)$ and the 
definition of $\apre(\cdot,\cdot)$ and $\spre(\cdot)$ it follows that
for all $i>0$, given $\rank(i-1)=X_i$, we have 
$\rank(i)=X_{i+1} = \apre(Q_\as^\equi,X_i) \lor (B_{\target} \land \spre(Q_\as^\equi))$.
By induction we have 
\[Q^*=\sem{\mu X.  \big(\apre(Q_\as^\equi,X) \lor 
(B_{\target} \land \spre(Q_\as^\equi))\big)}\,.\]
Since $Q^*=Q_\as^\equi$ (by Lemma~\ref{lemm:randomized3})  we obtain
the desired result.
Since $Z$ is the greatest fixed-point we have $Q_\as^\equi \subseteq Z$.
\end{enumerate}
The result follows.\qed

\begin{thm}[Complexity of almost winning for B\"uchi objectives 
under imperfect information]\label{thrm:randomized3}
Let $G$ be a game structure of imperfect information, let $\target$ be a 
set of observations, and let $\ell$ be a state of $G$.
Whether $\l$ is an almost-winning state in $G$ for 
the B\"uchi objective $\Buchi(\target)$ can be decided in {\sc Exptime}.
\end{thm}

The facts that $Z=Q_\as^\equi$ and that $H$ is exponential in the
size of $G$ yield Theorem~\ref{thrm:randomized3}.
The arguments for the proofs of Theorem~\ref{thrm:randomized2}
and~\ref{thrm:randomized3} do not directly extend to coB\"uchi 
or parity objectives.
In fact, Theorem~\ref{thrm:randomized2} does not hold for parity objectives 
in general, for the following reason: 
in concurrent games with parity objectives
with more than two priorities, almost-winning strategies may require 
infinite memory; for an example, see~\cite{dAH00}.
Such concurrent games are reducible to semiperfect-information 
games~\cite{ChatterjeeH05}, and semiperfect-information games are reducible 
to the imperfect-information games we study.
Hence a reduction to finite game structures of perfect information in order 
to obtain randomized positional strategies is not possible with respect to 
almost winning for general parity objectives.
Theorem~\ref{thrm:randomized2} and Theorem~\ref{thrm:randomized3}  may hold
for coB\"uchi objectives, but there does not seem to be a simple extension 
of our arguments for B\"uchi objectives to the coB\"uchi case.
The results that correspond to Theorems~\ref{thrm:randomized2} 
and~\ref{thrm:randomized3} for coB\"uchi objectives are open.

\medskip\noindent{\em Direct symbolic algorithm.} 
As in Section~\ref{sec:interpretations-mu-calculus}, 
the subset structure $H$ does not have to be constructed explicitly.
Instead, we can evaluate a fixed-point formula on a
well-chosen lattice.
The fixed-point formula to compute the set $Q_\as^\equi$ is evaluated on 
the lattice $\tuple{2^{Q}, \subseteq, \cup, \cap, Q, \emptyset}$. It is 
easy to show that the sets computed by the fixed-point algorithm are downward
closed for the following order on $Q$: for $(s,\l),(s',\l') \in Q$, let 
$(s,\l) \preceq (s',\l')$ iff $\l = \l'$ and $s \subseteq s'$. Then, we 
can define an antichain over $Q$ as a set of pairwise 
$\preceq$-incomparable elements of $Q$, and compute the almost-sure winning 
states in the lattice of antichains over $Q$, without explicitly 
constructing the exponential game structure~$H$.


\section{Lower Bounds}

We show that deciding the existence of a deterministic (resp.\
randomized) observation-based sure-winning (resp.\ almost-winning)
strategy for Player~$1$ in games of imperfect information is
{\sc Exptime}-hard already for reachability objectives. 
A first proof for sure-winning was given in~\cite{Reif84}. We give all 
the details of the reduction used in the proof and show that it 
extends to almost winning as well.

\medskip\noindent{\em Sure winning.} To show the lower bound result,
we use a reduction of the membership problem for polynomial space
Alternating Turing Machine.
An \emph{alternating Turing machine} (ATM) is a tuple
$M = \tuple{Q, q_0, g, \Sigma_i,\Sigma_t,\delta, F}$ where:
\begin{enumerate}[$\bullet$]
\item 
$Q$ is a finite set of control states;
\item 
$q_0 \in Q$ is the initial state;
\item 
$g: Q \to \{\land, \lor\}$; 
\item 
$\Sigma_i= \{0,1\}$ is the input alphabet;
\item 
$\Sigma_t= \{0,1,2\}$ is the tape alphabet and $2$ is the \emph{blank} symbol;
\item 
$\delta \subseteq Q \times \Sigma_t \times Q \times \Sigma_t \times \{-1,1\}$ is a transition relation; and 
\item 
$F \subseteq Q$ is the set of accepting states.
\end{enumerate}
We say that $M$ is a \emph{polynomial space} ATM if for some
polynomial $p(\cdot)$, the space used by $M$ on any input word $w$ is
bounded by $p(\abs{w})$. 

Without loss of generality, we make the
hypothesis that the initial control state of the machine is a
$\lor$-state and that transitions link $\lor$-state to $\land$-state
and vice versa. A word $w$ is \emph{accepted} by an ATM $M$
if there exists a run tree of $M$ on $w$ whose all leaf nodes
are accepting configurations (see~\cite{ChandraKS81} for details).
The AND-OR graph of the polynomial space ATM $(M,p)$ on the input word $w \in \Sigma^*$ is 
$G(M,p) = \tuple{S_{\lor}, S_{\land}, s_0, \Rightarrow, R}$ where
\begin{enumerate}[$\bullet$]
\item $S_{\lor} = \{(q,h,t) \mid q \in Q,\; g(q) = \lor, 1 \leq h \leq p(\abs{w})$ and $t \in \Sigma_t^{p(\abs{w})}\}$;
\item $S_{\land} = \{(q,h,t) \mid q \in Q,\; g(q) = \land, 1 \leq h \leq p(\abs{w})$ and $t \in \Sigma_t^{p(\abs{w})}\}$;
\item $s_0 = (q_0, 1, t)$ where $t = w.\Sigma_t^{p(\abs{w})-\abs{w}}$;
\item $((q_1,h_1,t_1),(q_2,h_2,t_2)) \in \Rightarrow$ iff there exists $(q_1, t_1(h_1), q, \gamma,d) \in \delta$ such that
$q_2 = q$, $h_2 = h_1 + d$, $t_2(h_1) = \gamma$ and $t_2(i) = t_1(i)$ for all $i \neq h_1$;
\item $R = \{(q,h,t) \in S_{\lor} \cup S_{\land} \mid q \in F \}$.
\end{enumerate}
A word $w$ is \emph{accepted} by $(M,p)$ iff $R$ is reachable in $G(M,p)$. 
The \emph{membership problem} is to decide if a given word $w$ is
accepted by a given polynomial space ATM $(M,p)$. This problem is
known to be {\sc ExpTime-hard}~\cite{ChandraKS81}.

\medskip\noindent{\em Idea of the reduction.}
Given a polynomial space ATM $M$ and a word $w$, we construct a game 
of size polynomial in the size of $(M,w)$ to simulate the execution 
of~$M$ on~$w$. Player~$1$ makes choices in $\lor$-states and Player~$2$ makes
choices in $\land$-states.  Furthermore, Player~$1$ is responsible for
maintaining the symbol under the tape head. The objective is to reach
an accepting configuration of the ATM.

Each turn proceeds as follows. In an $\lor$-state,
by choosing a letter $(t,a)$ in the alphabet of the game, Player~$1$ reveals 
$(i)$ the transition $t$ of the ATM that he has chosen
(this way he also reveals the symbol that is currently under the tape head)
and $(ii)$ the symbol $a$ under the next position of the tape head.
If Player~$1$ lies about the current or the next symbol under the tape head, he
should loose the game, otherwise the game proceeds. The machine is now
in an $\land$-state and Player~$1$ has no choice: he announces a special 
symbol $\epsilon$ and Player~$2$, by resolving nondeterminism on $\epsilon$, 
chooses a transition of the Turing machine which is compatible with the current
symbol under the tape head revealed by Player~$1$ at the previous turn.
The state of the ATM is updated and the game proceeds.  The transition
chosen by Player~$2$ is visible in the next state of the game and
so Player~$1$ can update his knowledge about the configuration of the ATM. 
Player~$1$ wins whenever an accepting configuration of the ATM is reached,
that is $w$ is accepted.

The difficulty is to ensure that Player~$1$ looses when he announces a
wrong content of the cell under the tape head.  As the number of
configurations of the polynomial ATM is exponential, we cannot
directly encode the full configuration of the ATM in the states of the
game. To overcome this difficulty, we use the power of imperfect
information as follows.  Initially, Player~$2$ chooses a position $k$,
$1 \leq k \leq p(|w|)$, on the tape: this number as well as the symbol
$\sigma \in \{0,1,2\}$ that lies in the tape cell number $k$ is
maintained all along the game in the non-observable portion of the
game states.  The pair $(\sigma,k)$ is thus private to Player~$2$ and
invisible to Player~$1$.  Hence, at any point in the game, Player~$2$
can check whether Player~$1$ is lying when announcing the content of
cell number $k$, and go to a sink state if Player~$1$ cheats (no other
states can be reached from there).  Since Player~$1$ does not know
which cell is monitored by Player~$2$ ($k$ is private), to avoid
loosing, he should not lie about any of the tape cells and thus he
should faithfully simulate the machine.
Then, he wins the game if and only if the ATM accepts the words $w$.

\medskip\noindent{\em Almost winning.}
To establish lower bound for almost-winning, we can use the same
reduction. Randomization can not help Player I in this game. Indeed,
at any point of the game, if Player I takes a chance in either not
faithfully simulating the ATM or lying about the symbol under the tape
head, the sink state is reached. In those case, the probability to
reach the sink state is positive and so the probability to win the
game is strictly less than one.
We now present the details of the reduction of the hardness proof.

\medskip\noindent{\em Reduction.}
Given a polynomial space ATM $(M,p)$, with $M=\tuple{Q, q_0, g,
  \Sigma_i,\Sigma_t,\delta, F}$ and a word $w$, we construct the
following game structure
$G_{M,p,w}=\tuple{L,\li,\Sigma,\trans,\Obs,\gamma}$, where:
  \begin{enumerate}[$\bullet$]
  \item The set of positions $L=\{ {\sf init} \} \cup \{ \sink \} \cup
    L_1 \cup L_2$ where: $L_1=(\delta \cup \{ - \}) \times Q \times \{
    1,\dots,p(|w|) \} \times \{ 1,\dots,p(|w|)\} \times \Sigma_t$. A
    state $(t,q,h,k,\sigma)$ consists of a transition $t\in \delta$ of the ATM
    chosen by Player~$2$ at the previous round or $-$ if this is the
    first round where Player~$1$ plays, the current control state $q$ of
    $M$, the position $h$ of the tape head,
    the pair $(k,\sigma)$ such that the $k$-th symbol of the tape is
    $\sigma$, this pair $(k,\sigma)$ will be kept invisible for Player~$1$.
    $L_2=Q \times \{ 1,\dots,p(|w|) \} \times \Sigma_t \times \{
    1,\dots,p(|w|) \} \times \Sigma_t$. A state
    $(q,h,\gamma,k,\sigma)$ consists of $q$, $h$, $k$, $\sigma$ as in
    $L_1$ and $\gamma$ is the symbol that Player~$1$ claims to be under
    the tape head.  The objective for Player~$1$ will be to reach a
    state $\l \in L$ associated with an accepting control state of $M$.
    \item $\li={\sf init}$. 
    \item $\Sigma=\{ \epsilon \} \cup ( \delta \times \Sigma_t )$.
    \item The transition relation $\trans$ contains the following sets of transitions:
      \begin{enumerate}[-]
      \item $I_1$ that contains transitions $({\sf
          init},\epsilon,(-,q_0,1,k,\sigma))$ where $(i)$ $1 \leq k
        \leq p(|w|)$ and $(ii)$ $\sigma=w(k)$ if $1 \leq k \leq |w|$
        and $\sigma=2$ otherwise. $I_2$ that contains transitions
        $({\sf init},(t,\gamma),{\sf sink})$ where $(t,\gamma) \in
        \delta \times \Sigma_t$.  So, at the initial state ${\sf init}$,
        Player~$1$ has to play $\epsilon$ in order to avoid entering
        ${\sf sink}$. By resolving nondeterminism on $\epsilon$,
        Player~$2$ chooses a tape cell to monitor.
      \item $S$ that contains transitions $({\sf sink},\sigma,{\sf
          sink})$ for all $\sigma \in \Sigma$. 
        When the ${\sf sink}$ state is entered, the game stays there forever.
      \item $L_{1.1}$ that contains transitions
        $(\l_1,\epsilon,{\sf sink})$ for all $\l_1 \in L_1$;  $L_{1.2}$ that
        contains transitions
        $((t,q,h,k,\sigma),((q_1,\gamma_1,q_2,\gamma_2,d),\gamma_3),{\sf
          sink})$ where $q_1 \not=q$ or $\neg ( 1 \leq h+d \leq
        p(|w|))$; $L_{1.3}$ contains the transitions
        $((t,q,h,k,\sigma),((q_1,\gamma_1,q_2,\gamma_2,d),\gamma_3),{\sf
          sink})$ where $h=k \land \gamma_1 \not= \sigma$ or $h+d=k
        \land \gamma_3 \not= \sigma$; $L_{1.4}$ contains the
        transitions
        $((t,q,h,k,\sigma),((q_1,\gamma_1,q_2,\gamma_2,d),\gamma_3),
        (q_2,h+d,\gamma_3,k,\sigma'))$ such that $q=q_1$, $1 \leq h+d
        \leq p(|w|)$, $h=k \rightarrow (\gamma_1=\sigma \land
        \sigma'=\gamma_2)$, and $h \not= k \rightarrow
        \sigma'=\sigma$.  Those transitions are associated with states
        of the game where Player~$1$ chooses a transition of the ATM to
        execute (if he proposes $\epsilon$, the game evolves to the
        ${\sf sink}$ state, see $L_{1.1}$). The transition proposed by
        Player~$1$ should be valid for the current control state of the
        ATM and the head should not exit the bounded tape after
        execution of the transition by the ATM, otherwise the game
        evolves to the ${\sf sink}$ state, see $L_{1.2}$. When
        choosing a letter, Player~$1$ also reveals the current letter
        under the tape head (given by the transition) as well as the
        letter under the next position of the tape head. If one of
        those positions is the one that is monitored by Player~$2$, the
        game evolves to the sink state in case Player~$1$ lies, see
        $L_{1.3},L_{1.4}$.
      \item $L_{2.1}$ contains the transitions
        $((q,h,\gamma_1,k,\sigma),\epsilon,
        ((q_1,\gamma_2,q_2,\gamma_3,d),q_3,h+d,k,\sigma'))$ such that
        $q=q_1$, $q_2=q_3$, $\gamma_1=\gamma_2$, $1 \leq h+d \leq
        p(|w|)$, $h=k \rightarrow \sigma'=\gamma_3$, and $h \not= k
        \rightarrow \sigma'=\sigma$; $L_{2.2}$ contains the
        transitions $((q,h,\gamma_1,k,\sigma),\epsilon,{\sf sink})$
        such that there does not exist a transition
        $(q,\gamma_1,q_1,\gamma_2,d) \in \delta$ with $1 \leq h+d \leq
        p(|w|)$;  $L_{2.3}$ contains the transitions
        $((q,h,\gamma_1,k,\sigma),(t,\gamma),{\sf sink})$ where
        $(t,\gamma) \in \Sigma \setminus \{ \epsilon \}$. Those
        transitions are associated with states of the game where
        Player~$2$ chooses the next transition of the ATM to execute.
        Player~$1$ should play $\epsilon$ otherwise the game goes to the
        ${\sf sink}$ state (see $L_{2.3}$). Also the game goes to the
        ${\sf sink}$ state if there is no valid transition to execute
        in the ATM (see $L_{2.2}$). In the other cases, when Player~$1$
        proposes $\epsilon$, Player~$2$ chooses a valid transition
        by resolving nondeterminism. The copy of the monitored cell is
        updated if necessary.
      \end{enumerate} 
    \item $\Obs=\{ {\sf init}, {\sf sink}\} \cup \Obs_1 \cup \Obs_2$
      where $\Obs_1=\{ (t,q,h) \mid \exists (t,q,h,k,\sigma) \in L_1
      \}$ and $\Obs_1=\{ (q,h,\gamma) \mid \exists (q,h,\gamma,k,\sigma) \in L_2 \}$.
  \item $\gamma$ is defined as follows: $\gamma({\sf
      init})=\{{\sf init}\}$, $\gamma({\sf sink})=\{{\sf sink}\}$, for
    all $(t,q,h) \in \Obs_1$, $\gamma(t,q,h)=\{ (t,q,h,k,\sigma) \in
    L_1 \}$, for all $(q,h,\gamma) \in \Obs_2$, $\gamma(q,h,\gamma)=\{
    (q,h,\gamma,k,\sigma) \in L_2 \}$.
  \end{enumerate}
  
  Finally, the objective $\phi$ of this game for Player~$1$ is to reach
  a state where the associated control state of the ATM is accepting,
  i.e. $\phi=\{ o_1 o_2 \dots o_n \dots \in \Obs^{\omega} \mid \exists
  i \geq 0 : (o_i=(t,q,h) \in \Obs_1 \lor o_i=(q,h,\gamma) \in \Obs_2)
  \land q\in F\}$.
  
  It follows that Player~$1$ has an observation-based sure-winning (or
  almost-winning) strategy in the game $G_{M,p,w}$ for the objective
  $\phi$ iff the word $w$ is accepted by the polynomial space ATM
  $(M,p)$. 
  This gives us Lemma~\ref{lemm:lower-bound} and 
  Theorem~\ref{thrm:lower-bound} follows from the lemma.

\begin{lem}\label{lemm:lower-bound} 
  Player~$1$ has a deterministic (resp. randomized) observation-based
  sure-winning (resp. almost-winning) strategy in the game $G_{M,p,w}$
  for the objective $\phi$ iff the word $w$ is accepted by the
  polynomial space ATM $(M,p)$.
\end{lem}

\begin{thm}[Lower bounds]\label{thrm:lower-bound}
Let $G$ be a game structure of imperfect information, let $\target$ be 
a set of observations, and let $\l$ be a state of $G$.
Deciding whether $\l$ is a sure-winning state in $G$ for the reachability 
objective $\Reach(\target)$ is {\sc Exptime}-hard.
Deciding whether $\l$ is an almost-winning state in $G$ for $\Reach(\target)$ 
is also {\sc Exptime}-hard.
\end{thm}

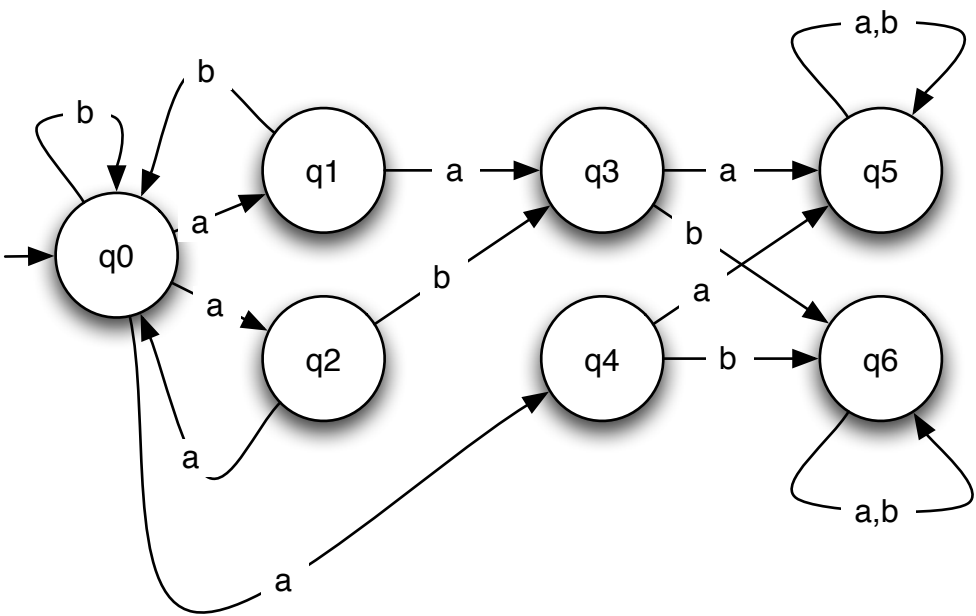
\end{document}

%% file: figure-example.tex
\unitlength=.6mm
\def\fsize{\scriptsize}

\begin{picture}(118,52)(0,10)

{\fsize
\gasset{Nw=9,Nh=9,Nmr=4.5,rdist=1,Nadjust=n}

\node[Nmarks=i,NLangle=0.0](q0)(15,29){{\fsize $\l_1$}}
\node[Nmarks=n](q1)(45,44){$\l_2$}
\node[Nmarks=n](q2)(45,14){$\l'_2$}
\node[Nmarks=n](q3)(75,44){$\l_3$}
\node[Nmarks=n](q4)(75,14){$\l'_3$}
\node[Nmarks=r](q5)(105,14){$\l_4$}

\drawloop[ELside=l,loopCW=y, loopangle=0, loopdiam=6](q5){$a,b$}

\drawedge[ELpos=50,ELside=l](q0,q1){$a,b$}
\drawedge[ELpos=50,ELside=r](q0,q2){$a,b$}
\drawedge(q1,q3){$a$}
\drawedge[ELside=r](q2,q4){$a$}
\drawedge[ELpos=32, ELside=l](q1,q4){$b$}
\drawedge[ELpos=32, ELside=r](q2,q3){$b$}
\drawbpedge[ELpos=78, ELside=r, ELdist=1, syo=3, exo=-1](q3,165,20,q0,80,50){$a,b$}
\drawedge[ELpos=50](q4,q5){$a,b$}

\gasset{Nmr=0,Nframe=y,Nadjust=n,dash={1.5}0,AHnb=0}

\node[Nw=18,Nh=15](c1)(15,29){}
\node[Nw=18,Nh=45](c2)(45,29){}
\node[Nw=18,Nh=45](c3)(75,29){}
\node[Nw=18,Nh=15](c4)(105,14){}

\gasset{Nframe=n}

\node(o1)(15,3){$o_1$}
\node(o2)(45,3){$o_2$}
\node(o3)(75,3){$o_3$}
\node(o4)(105,3){$o_4$}

}
\end{picture}

%% file: ex-rand-gastex.tex
\unitlength=.6mm
\def\fsize{\scriptsize}

\begin{picture}(120,70)(0,0)

{\fsize
\gasset{Nw=9,Nh=9,Nmr=4.5,rdist=1,Nadjust=n}

\node[Nmarks=i,NLangle=0.0](l0)(15,35){$\l_0$}
\node(l1)(45,50){$\l_1$}
\node(l2)(45,20){$\l_2$}
\node(l3)(75,50){$\l_3$}
\node(l4)(75,20){$\l_4$}
\node[Nmarks=r](l5)(105,50){$\l_5$}
\node(l6)(105,20){$\l_6$}

\drawloop[ELside=l,loopCW=y, loopdiam=6](l0){$b$}
\drawloop[ELside=l,loopCW=y, loopdiam=6](l5){$a,b$}
\drawloop[ELside=r,loopCW=n, loopangle=-90, loopdiam=6](l6){$a,b$}

\drawedge[ELside=l,ELpos=50, syo=2,curvedepth=-6](l0,l1){$a$}
\drawedge[ELside=r,ELpos=50, syo=-2,curvedepth=6](l0,l2){$a$}
\drawbpedge[ELside=r, ELpos=45, curvedepth=16](l0,-80,25,l4,-120,35){$a$}
\drawedge[exo=2, ELside=r, curvedepth=-6](l1,l0){$b$}
\drawedge[curvedepth=6](l2,l0){$a$}
\drawedge(l1,l3){$a$}
\drawedge[ELside=r, ELpos=48](l2,l3){$b$}
\drawedge(l3,l5){$a$}
\drawedge[ELpos=25, ELside=r](l3,l6){$b$}
\drawedge[ELpos=25](l4,l5){$b$}
\drawedge[ELside=r](l4,l6){$a$}

\gasset{Nmr=0,Nframe=y,Nadjust=n,dash={1.5}0,AHnb=0}

\node[Nw=18,Nh=15](c1)(15,35){}
\node[Nw=18,Nh=45](c2)(45,35){}
\node[Nw=18,Nh=45](c3)(75,35){}
\node[Nw=18,Nh=15](c4)(105,50){}
\node[Nw=18,Nh=15](c4)(105,20){}

\gasset{Nframe=n}

\node(o1)(15,24){$o_1$}
\node(o2)(45,61){$o_2$}
\node(o3)(75,61){$o_3$}
\node(o4)(105,39){$o_4$}
\node(o4)(105,31){$o_5$}

}
\end{picture}

%% file: reduction-gastex.tex
\unitlength=.6mm
\def\fsize{\scriptsize}

\begin{picture}(125,52)(0,0)

{\fsize
\gasset{Nw=9,Nh=9,Nmr=4.5,rdist=1,Nadjust=w}

\node[Nframe=n,ExtNL=y, NLdist=2, NLangle=-90](q0)(13,21){$\rank(3)$}
\node[Nmarks=i,NLangle=0.0](q0)(13,21){$\{\l_1\}, \l_1$}   

\node[Nframe=n,ExtNL=y, NLdist=1.5](q1)(43,36){$\rank(2)$}
\node[Nmarks=n](q1)(43,36){$\{\l_2,\l_2'\},\l_2$}

\node[Nframe=n,ExtNL=y, NLdist=2, NLangle=-90](q2)(43,6){$\rank(2)$}
\node[Nmarks=n](q2)(43,6){$\{\l_2,\l_2'\},\l_2'$}

\node[Nframe=n,ExtNL=y, NLdist=1.5](q3)(78,36){$\rank(4)$}
\node[Nmarks=n](q3)(78,36){$\{\l_3,\l_3'\},\l_3$}

\node[Nframe=n,ExtNL=y, NLdist=2, NLangle=-90](q4)(78,6){$\rank(1)$}
\node[Nmarks=n](q4)(78,6){$\{\l_3,\l_3'\},\l_3'$}

\node[Nframe=n,ExtNL=y, NLdist=2, NLangle=-90](q6)(113,6){$\rank(0)$}
\node[Nmarks=r, Nw=16, Nadjust=w](q6)(113,6){$\{\l_4\},\l_4$}

\drawloop[ELside=l,loopCW=y, loopangle=90, loopdiam=8](q6){$a,b$}

\drawedge[ELpos=46, ELdist=0.5, exo=-5, eyo=2](q0,q1){$a,b$}
\drawedge[ELpos=50, ELdist=0.5, exo=-5, eyo=2](q0,q2){$a,b$}
\drawedge[ELside=l, ELpos=48](q1,q3){$a$}
\drawedge[ELside=r, ELpos=48](q2,q4){$a$}
\drawedge[ELpos=32, ELside=l, ELdist=1, sxo=3, exo=-3](q1,q4){$b$}
\drawedge[ELpos=32, ELside=r, ELdist=0.5, sxo=3, exo=-3](q2,q3){$b$}
\drawedge[ELpos=70, ELside=r,curvedepth=-24, exo=-3](q3,q0){$a,b$}
\drawedge[ELpos=53](q4,q6){$a,b$}



\gasset{Nframe=n,dash={1}0,AHnb=0, Nadjust=wh}

\node[Nframe=n](q12)(43,21){$\equi$}
\node[Nframe=n](q34)(78,21){$\equi$}
\drawedge[AHnb=0](q1,q12){}
\drawedge[AHnb=0](q12,q2){}
\drawedge[AHnb=0](q3,q34){}
\drawedge[AHnb=0](q34,q4){}


}
\end{picture}